# Massive MIMO Evolution Towards 3GPP Release 18

Huangping Jin, Kunpeng Liu, Gilwon Lee, Emad J. Farag, Min Zhang, Dalin Zhu, Leiming Zhang,
Eko Onggosanusi, Mansoor Shafi, *Life Fellow*, *IEEE*, and Harsh Tataria, *Member, IEEE*

*Abstract*— Since the introduction of fifth-generation new radio (5G-NR) in Third Generation Partnership Project (3GPP) Release 15, swift progress has been made to evolve 5G with 3GPP Release 18 emerging. A critical aspect is the design of massive multiple-input multiple-output (MIMO) technology. In this line, this paper makes several important contributions: We provide a comprehensive overview of the evolution of standardized massive MIMO features from 3GPP Release 15 to 17 for both time/frequency-division duplex operation across bands FR-1 and FR-2. We analyze the progress on channel state information (CSI) frameworks, beam management frameworks and present enhancements for uplink CSI. We shed light on emerging 3GPP Release 18 problems requiring imminent attention. These include advanced codebook design and sounding reference signal design for coherent joint transmission (CJT) with multiple transmission/reception points (multi-TRPs). We discuss advancements in uplink demodulation reference signal design, enhancements for mobility to provide accurate CSI estimates, and unified transmission configuration indicator framework tailored for FR-2 bands. For each concept, we provide system level simulation results to highlight their performance benefits. Via field trials in an outdoor environment at Shanghai Jiaotong University, we demonstrate the gains of multi-TRP CJT relative to single TRP at 3.7 GHz.

*Index Terms*—3GPP, 5G-NR, beamforming, codebooks, coherent joint transmission, CSI, and massive MIMO.

## I. INTRODUCTION

THE utilization of massive multiple-input multiple-output (MIMO) technology has been one of the defining features of fifth-generation new radio (5G-NR) systems [1,2]. With the advent of active antenna arrays, configuring large numbers of antenna elements (tens to hundreds) at radio base stations (BSs) has become feasible from a design and implementation standpoint. Commonly, uniform planar arrays (a.k.a. uniform rectangular arrays) spanning two-dimensions (2D) are employed at BSs, to serve multiple users (UEs) within the same time-frequency resource via *spatial beamforming* across the azimuth and elevation domains [3]. Beamforming allows the BS to form multiple electronically steerable beams in the desired direction(s) of UE(s) to enhance coverage and spectral efficiency performance. Irrespective of the frequency band of operation, to ensure such higher performance in the downlink, 5G-NR systems use *link adaptation* to select a suitable modulation and coding rate (MCS) accommodating for the time-varying propagation channel conditions between the BS and UE(s) [4]. With time-division duplex (TDD) operation, the downlink transmission is adapted based on uplink channel estimates, assuming *reciprocity* between the uplink and downlink channels, respectively. In contrast, for frequency-division duplex (FDD) operation, link adaptation is based on the uplink *feedback* subsequent to the channel estimation in the downlink. The estimation and acquisition of channel knowledge at both the BS and UEs is referred to as *channel state information (CSI)* [5]. For both TDD and FDD operation, a CSI report in 5G-NR primarily comprises of three components: (1) *rank indicator (RI)*, (2) *precoding matrix identifier (PMI)*, and (3) *channel quality indicator (CQI)* [6].

The required massive MIMO features and functionality starkly differ between the two Third Generation Partnership Project (3GPP) classified frequency ranges (FR), i.e., FR-1 which spans from 410 to 7,125 MHz and FR-2 which spans from 24,250 to 52,600 MHz, respectively [2,5,7][1]. For systems operating in the FR-1 band, *digital beamforming* is typically implemented at the BS, where a large array of $N_\mathrm{t}$ dual-polarized transmit antennas is used to generate $N_\mathrm{b}$ digitally controlled beams ($N_\mathrm{b} \ll N_\mathrm{t}$) at the array baseband (prior to up-conversion). Here, each beam constitutes a *reference signal (RS)* [4]. Contrary to this, in the FR-2 band, almost all current commercial BSs only support *analog beamforming*, generating $N_\mathrm{b}$ analog beams (after up-conversion) from the available $N_\mathrm{t}$ transmit antennas, each constituting a RS resource. FR-2 bands are primarily restricted to analog beamforming due to the implementation complexity and energy consumption issues arising from supporting wider carrier bandwidths at much higher center frequencies relative to FR-1 bands. In particular, lower power amplifier (PA) efficiencies (typically 10 to 20%) and the need for a dedicated digital-to-analog converter(s) (DAC(s)) for each antenna element limits the scope of FR-2 band transmission to localized *per-antenna* phase shifts with complimentary power divider and combiner circuits [7].

For FR-1 bands, the first 5G-NR specification, i.e., 3GPP Release 15, supports low (*Type-I*) and high (*Type-II*) resolution

This manuscript has been submitted on October 15, 2022. H. Jin, K. Liu, M. Zhang, and L. Zhang are with Huawei Technologies (e-mail: {jinhuangping, kunpeng.liu, min.zhang1,zhangleiming.zhang}@huawei.com). G. Lee, E. J. Farag, D. Zhu, and E. Onggosanusi are with Samsung Research America (e-mail: {gilwon.lee,e.farag,dalin.zhu,eko.o}@samsung.com). M. Shafi is with Spark New Zealand (e-mail: mansoor.shafi@spark.co.nz). H. Tataria is with Tataria Consulting (e-mail: harshtataria@gmail.com).

---

[1] Typically, the state-of-the-art literature denotes the FR-1 band as conventional *microwave* bands, while the FR-2 bands are referred to as *millimeter-wave (mmWave)* bands, see e.g., [2] and references therein.

codebooks (a.k.a. *Type-I and Type-II CSI*) for beamforming to single and multiple UEs [8]. For multiuser scenarios, Type-II codebooks are intended to facilitate quantized channel eigenvector feedback, which is especially beneficial for combating inter-UE interference while spatially multiplexing data streams to multiple UEs. 3GPP Release 16 built on the codebooks introduced in Release 15 and introduced enhanced Type-II codebooks which reduced the associated feedback overheads via the use of spatial domain compression, while facilitating transmission of 4 spatial streams per-UE relative to 2 streams in Release 15 [9]. Further enhancements were made in 3GPP Release 17 to reduce the CSI overheads by exploiting (partial) angle-delay reciprocity in the propagation channel, while extending the applicable scenarios to incorporate UE mobility beyond pedestrian levels [10]. For downlink transmission in the FR-2 bands, CSI-RS and synchronization signal blocks (SSBs) are used, where a signal transmitted along a beam is received at the UE(s) by selecting *one* of the candidate receive beams that best matches the transmitted beam in terms of the received signal-to-noise ratio (SNR) of the UE(s). A host of 3GPP procedures to facilitate this operation is referred to as *beam management* (see [11]). While 3GPP Release 15 specifies beam management sufficient to support fixed wireless access services, its overheads and latency become a bottleneck for moderate-to-high-speed UE mobility, which is addressed by 3GPP Release 16 and 17, respectively [12]. Additionally, Release 15 lacks support for carrier aggregation, which is alleviated in Release 16 and 17. For both FR-1 and FR-2 bands, CSI enhancements for the uplink are also heavily investigated with proposals for different codebook vs. non-codebook approaches [13] and the possibility of utilizing different downlink modulation reference signal port (DMRS) configuration [14].

Despite the above advances, to the best of our knowledge, the existing literature does not *holistically* capture most aspects of massive MIMO evolution for both TDD and FDD systems, across both FR-1 and FR-2 bands from the beginning of Release 15 to Release 18. We bridge this gap by addressing different aspects of massive MIMO evolution in a continuum.

In parallel to the above developments, 3GPP Release 16 and 17 standardized the so-called multi-transmission/reception point (multi-TRP) functionality [15]. This helped to increase the *reliability* of 5G-NR systems by providing spatial macro diversity so that if one propagation path is blocked, an alternative path from a second TRP can be used for data transmission. In particular, the *transmission configuration indicator* (TCI) state was designed to play a key role in multi-TRP transmission, providing information required to identify and track a reference signal for downlink reception. In the uplink, a counterpart of TCI state is *spatial relation*, which identifies a reference signal to be used as an uplink beam, especially for FR-2 bands. In the case of multi-TRP transmission, more than one TCI state or spatial relation is signaled to UE for transmission/reception of a given channel.

The 3GPP Release 16 multi-TRP transmission mechanisms support both *coherent and non-coherent joint transmission, a.k.a., CJT and NCJT*, for TDD and FDD systems, respectively [15]. This results in significant downlink data rate increases (particularly for UEs that are closer to the radio cell-edge) alongside an increase in system reliability. The studies in [16-18] discuss the progress made in 3GPP Release 16 and 17 standardization of CJT and NCJT for both TDD and FDD multi-TRP systems for both FR-1 and FR-2 bands.

Nonetheless, for CJT systems operating in the FR-1 band under FDD mode, *optimal* design of beamforming codebooks from a CSI overhead minimization and spectral efficiency maximization viewpoint remains an open problem. This is in addition to selecting an optimal *TRP coordination set* for enhancing the overall system performance for joint transmission. To address this gap, we propose a modification to the contemporaneous 3GPP Release 16 and 17 oversampled and orthogonal 2D discrete Fourier transform (DFT) beams used in codebook design, where we incorporate channel feedback based on spatial and frequency domain basis. Our idea is to utilize the *channel correlation* in either frequency or spatial domain to seek a *sparse basis* for an approximate representation of the overall beamforming matrix which helps to reduce the required overheads and helps to improve performance. Likewise, For CJT systems operating in the TDD mode for FR-1 bands, the strict requirements on the CSI accuracy drives the need for new solutions. This is since the beamformers designed at the remote radio heads (RRHs) of the coordinated cells are required to ensure that the *phase* of the received signals from different cells add up *constructively* based on the acquired CSI. In such a case, an important factor that limits the CSI accuracy is the *inter-cell interference*. The current interference suppression methods rely on whitening the colored interference by low correlation sequences during the least squares (LS) operation at the BS. However, the sequence correlation is limited by its length and is difficult to further improve. In such cases, to achieve better SNR for CSI acquisition, we propose a modification to the original CSI acquisition framework using sounding reference signals (SRS). Specifically, we extend the concept of *SRS frequency hopping*, leading to shorter root sequences being used. To resolve the inter-cell SRS interference/collision (especially for the case with limited SRS resources in the system), we introduce *SRS interference randomization* to improve the SRS-based channel estimation performance in scenarios with strong inter-cell interference. Via system-level simulations (SLS)[2], we later demonstrate the gains of using the abovementioned proposals for *both* FDD and TDD systems. To the best of our knowledge, this paper is the first to make multiple advances for CJT operation within the FR-1 bands.

In addition to the above, to demonstrate the gains achieved by CJT in a TDD system with single and multi-TRP, we present results from *a field measurement* conducted in an outdoor urban environment at Shanghai Jiaotong University, Shanghai, China.

---

[2] A SLS spans multiple layers of the open systems interconnection stack (physical, data link, access control) and implements 3GPP recommended procedures to compute UE or BS performance in a system setup as a network.

The measurements were carried out at 3.7 GHz center frequency across a bandwidth of 20 MHz, where both single and multi-UE capabilities were measured. The single UE measurement was carried out with a 5 kilometer/hour (km/h) UE mobility along a defined trajectory, whereas the multi-UE measurements focused on the performance enhancements delivered at the cell edge. Both cases compared the resulting UE spectral efficiency performance with multi-TRP relative to a single TRP case as a baseline. The transmission TRP set consisted of 2 TRPs (intra-site), while the interfering TRP set consisted of 4 TRPs (intra and inter-site). Each BS was equipped with 64 transmit/receive radio chains while each UE (Huawei Mate30) was equipped with 2 transmit and 4 receive radio chains. To the best of our knowledge, such an extensive measurement campaign has not previously been carried out for analyzing 3GPP compliant CJT, where the literature contains examples of measurement results in more simplified scenarios [19,20]. For FR-2 bands, the literature is sparse on the 3GPP Release 17 beam management enhancements in high mobility scenarios (channel coherence time proportional to UE velocities exceeding 100 km/h), see e.g., [6] and references therein. To bridge this gap, we present SLS in a dense urban highway (DUH) for beam management according to 3GPP Release 17 for the multi-TRP case. The motivation behind a unified TCI framework to reduce latency and overhead of beam indication is introduced, thereby enhancing system performance especially in high mobility scenarios. With a pre-defined path, the UE was designed to moving at a speed of 120 kmph (33.3 m/sec) and the UE's throughput was sampled every 1 m (30 ms), with 100 sample points i.e., total simulation duration is 3 seconds. SLS performance is evaluated for beam indication using downlink control information (DCI), with a latency of 0.5 ms and a BLER of 1% and using medium access control element (MAC-CE), with a latency of 3 ms and a BLER of 10%.

Collectively, the key contributions of this paper can be summarized as follows:

- We present a holistic overview of massive MIMO evolution in 3GPP standardization for both FR-1 and FR-2 frequency bands, across FDD and TDD configurations from Release 15 to 17. These form the basis for the current MIMO study item of Release 18, a.k.a. 5G-Advanced. Our overview includes the evolution of CSI acquisition frameworks and techniques; Enhancements towards higher-order massive MIMO support on the uplink via CSI and DMRS enhancements; CSI enhancements for mobility; and beam management frameworks. To the best of our knowledge, this is the first paper to address different aspects of massive MIMO evolution in a continuum.
- For FDD CJT systems with multi-TRPs, we consider the problem of optimal design of beamforming codebooks to minimize CSI overheads and maximize per-UE spectral efficiency. In this line, we propose a modification to the contemporaneous 3GPP Release 16 and 17 oversampled and orthogonal 2D DFT codebooks, where we incorporate channel feedback based on spatial and frequency domain basis. Utilizing the channel correlation in either frequency or spatial domain, we derive a sparse basis for an approximate representation of the overall beamforming matrix which helps to reduce the required overheads and helps to improve performance. Equivalently, for TDD-based CJT with multi-TRPs, we extend the concept of SRS frequency hopping and randomization, resolving the inter-cell SRS interference/collision while improving SRS-based channel estimation performance in scenarios with strong inter-cell interference. For both cases, we present SLS results for quantifying the resulting performance gains relative to a single TRP system.
- We present field measurements to demonstrate the gains achieved by CJT in a TDD system with single and multi-TRPs. The measurements were conducted in an outdoor urban environment at Shanghai Jiaotong University, Shanghai, China, at 3.7 GHz center frequency across a bandwidth of 20 MHz, where both single and multi-UE capabilities were measured. The single UE measurement was carried out with a 5 km/h UE mobility, whereas the multi-UE measurements considered four cases around on the performance enhancements delivered at the cell edges. Both cases compared the resulting UE spectral efficiency performance with multi-TRP relative to a single TRP case as a baseline. To the best of our knowledge, such an extensive measurement campaign has not been carried out for analyzing 3GPP compliant CJT previously.
- Via SLS, we demonstrate the enhancements on offer by 3GPP Release 17 beam management framework for FR-2 bands with high UE mobility. In particular, we show how a unified TCI framework reduces latency and overheads, improving system performance.

**Notation.** Upper and lower boldface letters represent matrices and vectors. The $M \times M$ identity matrix is denoted as $I_{M \times M}$. The Hermitian transpose and celling operations are denoted by $(.)^H$ and $\lceil . \rceil$, whereas the statistical expectation is denoted by $\mathbb{E}(.)$. Moreover, $\|.\|$ and $|.|$ denote the vector and scalar norms, while $\boldsymbol{a} \otimes \boldsymbol{b}$ denotes the Kronecker product of two vectors $\boldsymbol{a}$ and $\boldsymbol{b}$. Finally, we denote $O(1)$ as an order one term, while FFT(.) and IFFT(.) represent the fast Fourier transform and inverse fast Fourier transform, respectively. All other mathematical notation outside of what is stated here is defined within the text as necessary.

## II. MASSIVE MIMO TECHNOLOGY EVOLUTION IN 5G-NR

In this section, we discuss the evolution of massive MIMO technologies in 5G-NR systems, including the typical massive MIMO architecture at radio BSs, CSI reporting and acquisition frameworks for FDD and TDD systems, respectively, codebook-based beamforming, support in 3GPP for higher-order massive MIMO transmission/reception (serving many UEs), and mmWave enhancements in the FR-2 bands.

### A. Massive MIMO BS Architecture in 5G-NR

For both the FR-1 and FR-2 bands, *three categories* of BSs have been standardized by the 3GPP in TS 38.104; namely BS Type 1-*C*, 1-*H* and 1-*O*/2-*O* [21]. Most commercial FR-1 BSs are of type 1-*H*, where the radiation characteristics are defined over the air (OTA) and the conducted characteristics are defined at the boundary between the physical antenna ports (inclusive of radio distribution network) and transceiver antenna units. In contrast, commercial BSs for FR-2 bands are of the type 1-*O*/2-*O*, where the conducted characteristics are inclusive of the physical antenna ports *and* transceiver units. Taking the FR-1 bands as an example, the massive MIMO structure in terms of its architecture and design is depicted in Fig. 1. Such an architecture is common for 3GPP compliant commercial BSs and consists of three key components: (1) Dual-polarized micro-strip (patch) antenna elements; (2) Sub-array(s); and (3) An overall antenna array composing of multiple sub-arrays. In general, the deployed antenna elements are capable of simultaneously transmitting and receiving *two* mutually orthogonal signals leveraging maximum diversity in the polarization domain. Naturally, the element cross-polarization discrimination governs the spatial path isolation between the horizontal (H) and vertical (V) polarizations without increasing the form-factor of the element. Considering the fact that the azimuth and zenith angles would be confined to a range governed by the distribution of the UEs, and that FR-1 BSs are typically deployed on rooftops or masts, coverage across the full zenith range of -90° to 90° (usually) not required. To this end, a few adjacent antenna elements (typically in the vertical direction) are *grouped together* and are driven by a single up/down-conversion chain (including one PA and low-noise amplifier). This is known as a *sub-array*. An example of a 3×1 sub-array with elements spaced $d_v$ apart vertically in a massive MIMO radio is shown in Fig.1a). Since only *one* digital beamforming weight is applied within a given sub-array, the adjustment of a sub-array's steering angle requires an additional phase shifter network. The *overall array* then composes of multiple sub-arrays as its elements/components. As shown in Fig. 1a), four vertical columns of sub-arrays and four horizontal rows of sub-arrays create the 2D uniform planar array.

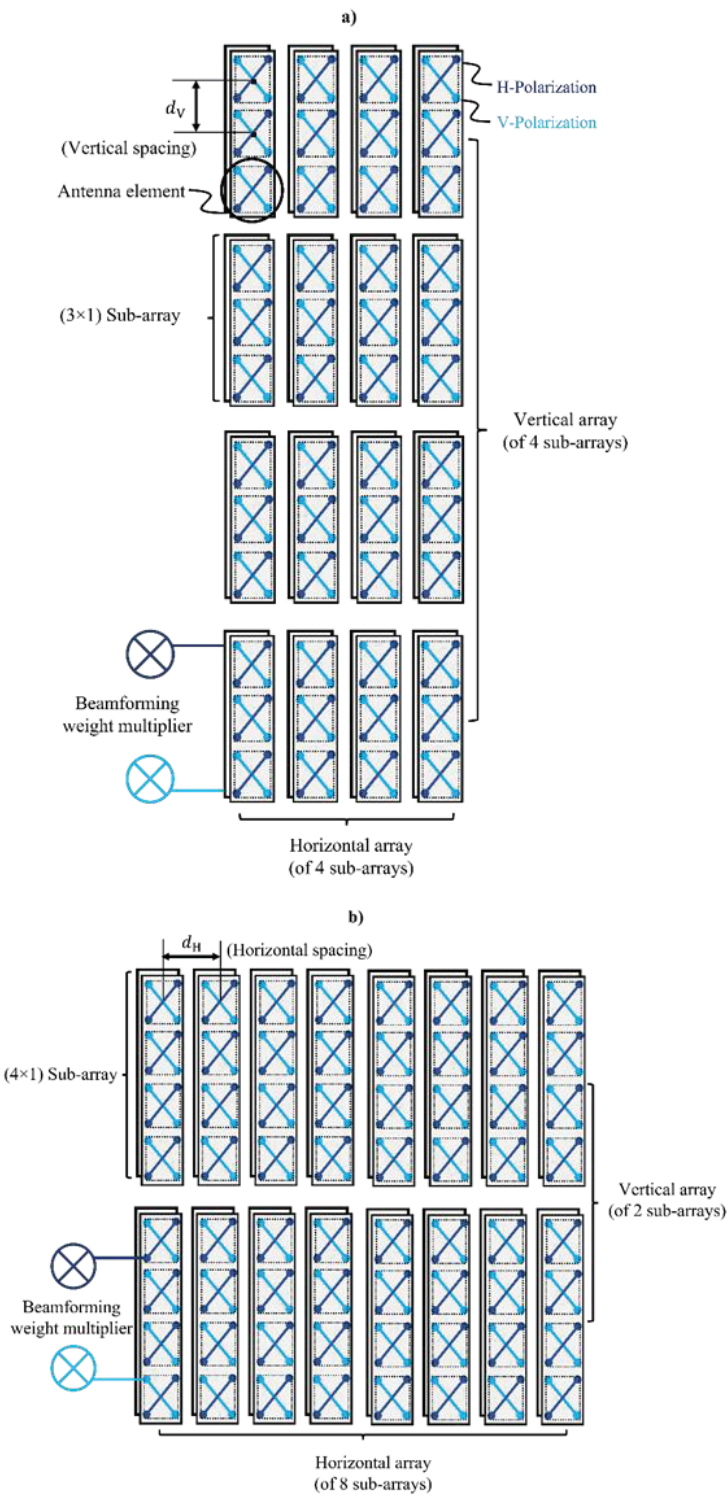

**Fig. 1.** a) and b): Two examples of a commercial massive MIMO architecture for FR-1 bands.

Such an architecture can be utilized by transmitting UE-specific beams towards individual receive antennas of the UE via a beamforming weight vector. Based on the three array components discussed above, the fundamental characteristics which determine 3GPP compliant commercial massive MIMO BS performance can be defined as follows:

- *UE-specific beam gain:* The maximum antenna gain of a UE-specific beam is expressed via a sum of its array gain and the sub-array gain, respectively. The array gain is proportional to the number of sub-arrays in a given array, and the sub-array gain is proportional to the number of elements grouped within a given sub-array, as well as the inter-element spacing between adjacent elements. The antenna gain towards the *array broadside* is governed by the antenna aperture of the massive MIMO BS, irrespective of its form factor.
- *Narrower beamwidths:* From array theory fundamentals, large arrays can yield narrower beams, which help to reduce the interference levels to other co-scheduled UEs who may be falling within the array's sidelobes. Naturally, the beamwidth is *inversely proportional* to the length of the massive MIMO array in both azimuth and elevation domains, respectively.
- Vertical angular coverage: The sub-array pattern does not change dynamically unless an additional phase shifter network is added. Thus, the operational range of the vertical steering angle is limited to the sub-array beamwidth – often referred to as the angular coverage of the massive MIMO system. Since a sub-array usually comprises of a single vertical column, as shown in Fig. 1 a) and b), the vertical angular coverage is inversely proportional to the length of the sub-array.

The above discussed massive MIMO architecture is used in 3GPP standardization to develop the necessary CSI acquisition and control/data signal transmission/reception frameworks. In what follows, we discuss some of these aspects, as well as their evolution across the different 3GPP standards releases from the beginning of Release 15 to Release 18.

### B. CSI Reporting and Acquisition for FDD and TDD Systems

The premise of beamforming, and more generally the evolution of massive MIMO in 3GPP standardization, is heavily dependent on the quality and accuracy of the acquired CSI at the BS and UEs. For FR-1 bands, FDD systems use CSI-RS as a signaling mechanism where the BS facilitates downlink transmission by sending a digitally beamformed signal towards the UE, which is received using digital combining across the UE's receive antennas. The BS derives its beamforming weights from CSI reports calculated by the UE according to a pre-defined codebook of beamforming vectors/matrices. In this line, 3GPP Release 15 supported Type-I (low) and Type-II (high-resolution) CSI codebooks, respectively [6]. Type-II CSI is intended to facilitate *refined resolution* quantized channel eigenvector feedback, which is tailored for spatially multiplexing data to multiple UEs, enabling higher-order massive MIMO operation. Equivalently, for downlink transmission in the FR-2 bands, CSI-RS and SSBs are utilized, where a signal transmitted along a beam is received by selecting *one* of the candidate receive beams that best matches the transmitted beam in terms of its received SNR. Here, 3GPP Release 15 has defined *beam management* procedures to facilitate this operation [22]. While these are sufficient to support fixed wireless access services, its overheads and latency become a bottleneck for moderate-to-high-speed UE mobility, which is addressed by Release 16 and 17, respectively. Additionally, Release 15 lacks support for carrier aggregation, which is alleviated in Release 16 and 17.

A typical CSI report consists of the following quantities:
- *RI:* This denotes what the *UE recognizes* as a suitable transmission rank, i.e., a suitable number of transmission layers, $L$, for downlink transmission. We note that this is independent of the scheduler decision, which will ultimately select the number of streams to transmit to a given UE based considering the RI feedback.
- *PMI:* Based on the selected rank, this metric indicates the beamforming matrix suitable to a given UE at a given time instance.
- *CQI:* This indicates a suitable channel coding rate and modulation scheme, given the selected precoder matrix. We note that this entry is utilized by the link adaption algorithms in 5G-NR.

*Each* possible value of PMI corresponds to *one* possible beamformer configuration. To this end, the set of all possible values of PMIs corresponds to a set of beamformers, referred to in the 3GPP terminology as *beamforming codebook,* which the UEs can select between when reporting PMI. In the sequel, we discuss the components and design of Type-I and Type-II codebooks, a.k.a., Type-I and II CSI, and present their evolution across the different 3GPP releases.

*1). Type-I Codebooks/CSI:* Type-I codebook/CSI primarily targets scenarios where a *single UE* is scheduled within a given time-frequency resource, potentially with transmission of a relatively large number of layers in parallel, i.e., higher-order spatial multiplexing. Type-I CSI composes of *two* sub-types of CSI, a.k.a., *Type-I Single Panel CSI* and *Type-I Multiple Panel CSI*, respectively, corresponding to different codebooks assuming single and multiple panel BS array configurations [13]. In the case for Type-I single panel CSI, the beamforming matrix, $W$, is then given by

$$W = W_1 W_2, \quad (1)$$

with information about the selected $W_1$ and $W_2$ reported separately as different parts of the PMI. Note that $W_1$ is designed to capture the *long-term frequency-independent* characteristics of the propagation channel. To this end, one $W_1$ is selected and reported for the *entire CSI reporting bandwidth*. In contrast, $W_2$ is designed to capture *short-term frequency-dependent* characteristics of the channel. Thus, $W_2$ is selected and reported on a *sub-band* basis, where a sub-band covers *part* of the overall CSI reporting bandwidth. In the case where the UE does not report $W_2$, the 5G-NR network selects $W_2$ on a per-physical resource block (PRB) basis. Mathematically, $W_1$ can be defined as a block diagonal matrix, describing a set of beams pointing to different directions, i.e.,

$$W_1 = \begin{bmatrix} B & 0 \\ 0 & B \end{bmatrix}, \quad (2)$$

where each column of the sub-matrices $B$ defines a beam and the 2×2 block diagonal structure is a result of supporting two-polarizations (H and V). Since $W_1$ is assumed to capture the long-term channel conditions, the same set of beams are assumed for both H and V-polarizations. Selecting $W_1$, and in turn selecting $B$, can be interpreted as selecting a (limited) set of beam directions from a large set of possible candidate directions defined by the full set of $W_1$ matrices within the codebook defined by the 3GPP. For the sake of saving overhead and complexity for reporting W2 which contributes the majority of overhead for a sub-band CSI reporting, either a single beam or four neighboring beams are defined, corresponding to four columns of the sub-matrix $B$ in the latter case. As for $W_2$, since it can be reported on a sub-band basis, it is possible to "fine tune" the beam directions per-sub-band. If $W_1$ only consists of a single beam, i.e., $B$ is a single column matrix, then $W_2$ only provides co-phasing between two polarizations. If $W_1$ defines $N$ neighboring orthogonal beams via techniques such as singular value decomposition (SVD) transmission, where the assumed $K \times NM$ propagation channel matrix, $H$, is decomposed into its SVD form $H = U \Lambda V^H$. For the sake of the argument, the number of antennas at the UE and BS are assumed to be $K$ and $NM$, respectively. From the SVD, $U$ is a $K \times K$ unitary matrix, $\Lambda$ is a $K \times K$ diagonal matrix of channel singular values, and $V$ is a $K \times NM$ unitary matrix such that $V^H V = I_{K \times K}$. Since the $\ell$-th diagonal element of $\Lambda$ encapsulates the $\ell$-th layer's channel magnitude, where $\ell \in \{1, 2, I, L\}$. The system can dynamically decide the number of layers to maximize massive MIMO spectral efficiency and select $v_l$, the $\ell$-th column vector of $V$ as the beamforming weight vector for the $\ell$-th layer. The 3GPP specifications support up *maximum of eight layers* to a single UE, i.e., $L=8$ in the above notation.

Due to the frequency selectivity, the channel response may be diverse at each PRB. Since CSI-RS can be transmitted over the entire bandwidth, each resource block can use different beamforming weights to maximize massive MIMO spectral efficiency. However, since one SVD operation in real-time requires highly complex calculations, executing SVDs for all PRBs is generally not implementation-friendly in terms of computational load deciding when to apply beamforming weights to downlink transmission. In light of this, simpler sub-optimal approaches may be used which are more implementation friendly.

In contrast to single-panel CSI, codebooks for Type-I multiple panel CSI are designed assuming the joint use of multiple antenna panels at the BS and caters for the fact that it may be difficult to ensure *coherence* between transmissions from different panels. The basic principle of Type-I multiple panel CSI is similar to that of Type-I single panel CSI. More specifically, the structure of $W_1$ is identical as for Type-I single-panel CSI, however $W_2$ can now provide *per-sub-band co-phasing not only across the polarizations but also across the multiple panels*. The 3GPP Type-I multi-panel CSI supports spatial multiplexing with up to *four* layers.

*2) Type-II Codebooks/CSI:* Relative to Type-I, Type-II codebooks/CSI provides channel information with significantly higher spatial granularity. Primarily, targeting the multi-UE scenario, it was intentionally limited to a maximum of rank two in Release 15. This was later extended to rank four in Release 16. The mathematical structure of Type-II CSI follows (1) and (2) with $W_1$ defining a set of beams that is reported as part of the CSI. Since up to four beams may be reported, this corresponds to up to four columns in $B$. For each beam, $W_2$ is designed to provide an amplitude and phase control. We now discuss the evolution of Type-II CSI towards enabling multi-UE capabilities in Release 16.

*3) 3GPP Release 16 Type-II Enhanced Codebook:* The reporting of a relatively large number of combining coefficients on a per-sub-band basis yields large CSI reporting overheads for FDD massive MIMO systems. Hence the Enhanced Type-II codebook/CSI in Release 16 enables the utilization of *frequency domain correlation across the sub-bands* to reduce signaling overheads. While doing this, Release 16 offers a 2× improvement in the frequency domain granularity of PMI reporting. This is jointly achieved via the concept of *frequency-domain units*, where a unit corresponds to either a sub-band or half-a-sub-band *in-conjunction* with the compression operation. The BS is provided with a recommended beamformer for a frequency-domain unit, relative to one beamformer per-sub-band for Release 15 Type-II CSI. For a given layer, $\ell$, the Release 16 Type-II CSI can be expressed as

$$[w_\ell^{(0)} w_\ell^{(1)} \cdots w_\ell^{(F-1)}] = W_1(\widetilde{W}_{2,\ell} W_{f,\ell}^H), \quad (3)$$

Where $f \in F$ is the total number of frequency-domain units to be reported. Note that the left-hand side of (3) describes the set of beamforming vectors for the *complete set* of frequency-domain units for a given layer. For a total of $L$ layers, there are $L$ such beamforming vectors, each consisting of $F$ different configurations. The actual reported beamformer which maps the involved massive MIMO layers to antenna ports for a given frequency-domain unit, $x$, can then be computed as

$$W^{(x)} = [w_0^{(x)} w_1^{(x)} \cdots w_{L-1}^{(x)}]. \quad (4)$$

The structure of $W_1$ is as given in (2) and is the same for all frequency-domain units and layers, respectively. The size of the compression matrix $W_{f,\ell}^H$ is $Z \times F$, which composes of a set of row vectors from a DFT basis and *provides a transformation from the frequency-domain of dimension $F$ into a smaller delay domain of dimension $Z$ ($Z < F$)*. We note that $W_{f,\ell}^H$ is frequency independent and is reported separately for each layer $\ell$. The number of rows of $W_{f,\ell}^H$ is equal to $Z=\lceil p(f/\varphi) \rceil$, where $\varphi$ is the number of frequency-domain units per-sub-frame and $p$ is a tunable parameter which controls the amount of compression. Furthermore, $\widetilde{W}_{2,\ell}$ maps a smaller version of delay domain to the beam domain, implying a smaller size of $\widetilde{W}_{2,\ell}$ and hence a smaller number of overall parameters to report. The overall concept of frequency-domain compression in Release 16 is depicted in Fig. 2. Since DFT vectors accurately approximate the eigenvectors of linear time-invariant and frequency-selective channels, the sparseness of PMI in frequency-domain (after DFT operation) allows PMI overhead reduction.

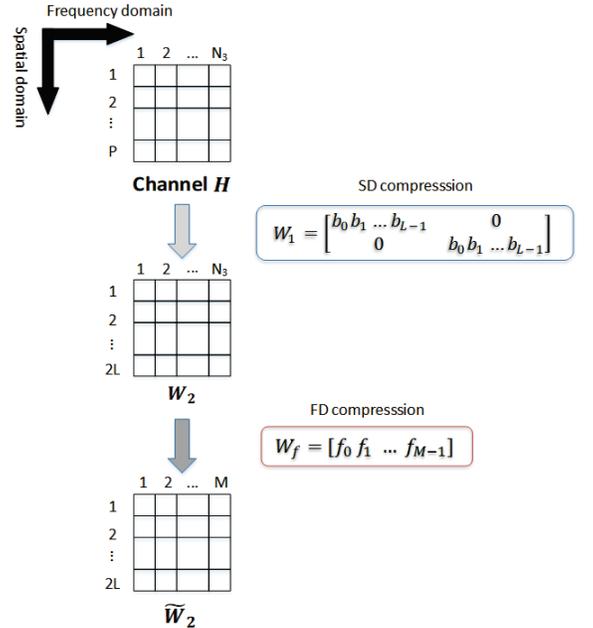

**Fig. 2.** Frequency-domain compression for Type-II CSI.

Although Type-II codebooks in Release 15 and 16 improve performance, both codebooks require a higher implementation complexity at the UE(s) for the sub-band PMI quantization. In addition, the *minimum* granularity of channel information must be equal to sub-band size or half sub-band size, which limits the CSI precision in frequency-domain. Furthermore, traditional DFT basis is used for angle and delay information quantization, by which the transformed spatial-frequency channel has lower sparsity than that by eigenvector of statistical spatial-frequency channel. To enable better trade-offs in UE complexity, CSI reporting overhead and performance, Release 17 Type-II CSI is introduced by utilizing *joint angle and delay reciprocity* between the downlink and uplink channels. Specifically, the BS will employ the angle-delay information obtained from the uplink channel to weight CSI-RS. To clearly explain the process outlined in Release 17, consider the following example: The BS estimates the angle and delay information corresponding to *three strongest angle-delay pairs* based on the uplink channel matrix, in which the information in frequency-domain is given by $f_1, f_2, f_3$ and the corresponding delays are denoted as $\tau_1, \tau_2$ and $\tau_3$. Then, the BS weights CSI-RS using these angle and delay pairs. By applying this frequency domain information on CSI-RS, it is equivalent to the BS shifting the delay of $p$-th angle-delay pair to a specific delay position, $\tau$. As a result, based on the CSI-RS weighted by angle and delay information, UE can obtain the coefficients according to the specific delay position. Followed by this, the UE can calculate the linear combination coefficient, $C_p$, for the $p$-th angle-delay pair on the CSI-RS for port $p$ based on a specific frequency information, avoiding the need to find the appropriate frequency-domain vectors.

For TDD systems, channel reciprocity holds between the downlink and uplink, and CSI acquisition is based on SRS initiated by the UE(s) towards the BS. Specifically, Release 15 defines the basic SRS functions for UEs equipped with up to 4 receive antennas. In practice, given that the number of receive radio chains at the UE is generally larger than the transmit radio chains, an *antenna switching* mechanism is supported. For instance, assuming that BS has 64 transmit/receive chains (a.k.a. 64T64R), and the UE has 2 transmit and 4 receive chains (a.k.a. 2T4R), via the principle of reciprocity, the BS needs CSI of the channel $H$ having dimensions of 4 x64 in order to design the downlink beamformer for the UE. The UE could sound the 4 receive chains across *two different times* to let the BS derive $H_1$ of the first two receive chains and $H_2$ of the remaining two receiving chains separately, where the dimension of both $H_1$ and $H_2$ is 2x64, and $H = ([H_1; H_2])^H$. As shown in Fig. 3, two orthogonal frequency-division multiplexing (OFDM) symbols are occupied by SRS from the UE, and each of the symbols contains the SRS transmitted via 2 of the 4 receive chains. Note that, each receive radio chain corresponds to an SRS port, and *different* SRS ports should thus occupy *orthogonal* physical resources.

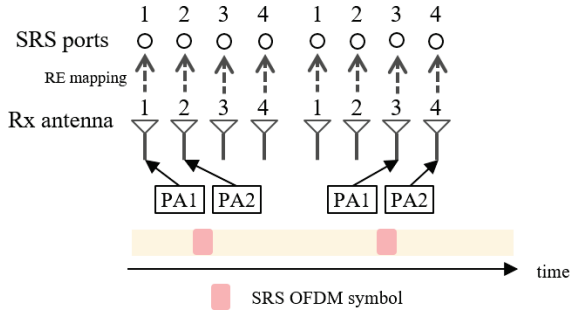

Fig. 3. Illustration of SRS antenna switching for downlink TDD CSI acquisition where the UE is equipped with a 2T4R architecture.

In contrast to Release 15, Release 16 introduced more flexible antenna switching mechanisms for SRS overhead reduction. For example, in Release 15, the 2T4R UE would be configured with 4 SRS ports, while in Release 16, the UE has the flexibility to fall back to 1T2R configuration, which means that only 2 SRS ports are needed to acquire CSI. The channel corresponding to the two remaining receive radio chains are not transmitted and can be derived using the *correlation* between the transmit antennas and the antennas that are not transmitting associated with the respective radio chains. Advancing further, Release 17 increases the antenna switching mechanism to 8 receive radio chains.

In addition to the above, Release 15 introduced the *SRS frequency hopping* mechanism for coverage enhancement. As shown in Fig. 4a), the UEs would transmit SRS on only *part* of the sounding bandwidth at each time and using *multiple times* to sound the whole system bandwidth. To this end, the power spectrum density is increased compared with the case when SRS is transmitted on whole band within one time slot. In Fig. 4a), the BS can estimate the CSI across the whole bandwidth after hopping 4 times. Release 17 further progressed the SRS spectral efficiency enhancement for frequency hopping. For instance, as shown in Fig. 4b), *partial SRS* is defined for each frequency hopping bandwidth, which allows each UE to transmit SRS only on a *part* of each frequency hopping bandwidth, while the BS could still reconstruct the full bandwidth channel. Since the re-use factor is 4 (Fig. 4b), 4 times the number of UEs can be supported relative to Release 15. This facilitates higher-order massive MIMO operation.

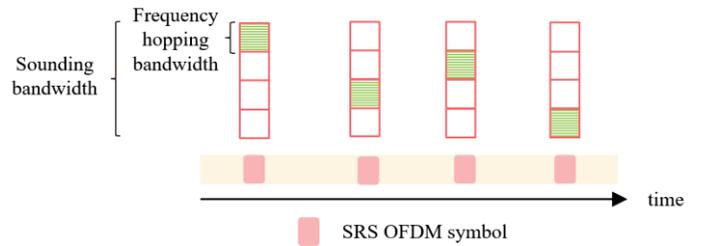

a) Illustration of SRS frequency hopping.

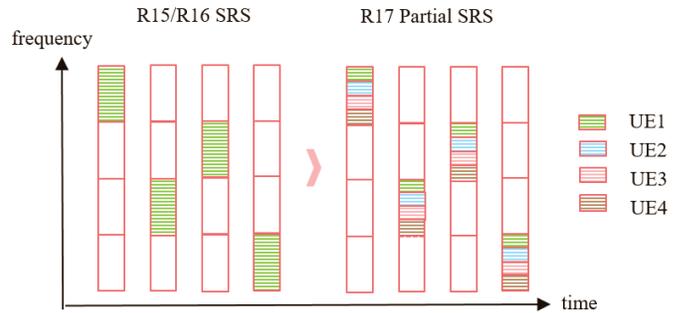

b) Evolution of SRS frequency hopping from Release 15/Release 16 to Release 17.

Fig. 4. Illustration of SRS transmission mechanism from 3GPP Release 15/Release 16 to Release 17.

*C. mmWave Applications*

In addition to the massive MIMO evolution aspects presented for the FR-1 band, a new key technology introduced in 5G-NR is the so-called *beam-based* operation. This is especially relevant for operation of systems in the FR-2 bands, i.e., above 24 GHz center frequency. As it is well known from fundamentals of wave propagation [23,24], higher frequencies suffer from high propagation loss. However, given the shorter wavelength, it is more feasible to pack more elements in a practical form-factor. Larger arrays provide narrower transmission and receptions beamwidths with higher gain to overcome the higher propagation loss. With a larger number of antenna elements, a fully digital transceiver (as in FR-1 bands) becomes prohibitive in terms of cost, size, and energy consumption, requiring analog-to-digital (ADC) and DAC converters for each antenna element. Instead, *analog* and *hybrid beamforming* architectures are adopted where a larger antenna array is *partitioned* in smaller number of radio frequency (RF) chains with analog phase-shifters networks to steer the beam in a desired direction. When transmitting, a combination of digital beamforming *before* the DAC and analog beamforming using the phase-shifters is used to create the *overall* beam shape.

3GPP Release 15 introduced *multi-beam operation*, where a beam refers to a spatial domain transmission filter or the spatial domain reception filter. Beam-based operation includes beam acquisition, beam maintenance and tracking, and beam failure recovery. During the *initial access* phase of a link, a UE identifies a SSB (which includes primary synchronization signal, secondary synchronization signal, and the physical broadcast channel (PBCH)) that is received with a reference signal received power (RSRP) that exceeds a pre-set threshold. Each SSB is transmitted using a corresponding spatial domain transmission filter. Thus, the SSB *selection* determines the beam to use for subsequent communication between the UE and BS. The UE selects a physical random-access channel (PRACH) occasion and a PRACH preamble that is associated with the SSB to use for transmission of the preamble, hence implicitly indicating the beam to the BS.

During the *beam maintenance* phase, for downlink beam indication and measurement, the reference signal can contain non-zero power channel state information reference signal (NZP CSI-RS) and/or SSB. Here, downlink beam indication is performed via TCI states. Each TCI state includes a TCI state identifier (ID) and quasi-co-location (QCL) information [12]. The QCL information includes a QCL type and a source RS. The QCL information associates one signal (e.g., data or control) with another (e.g., a source RS) that *shares* a same set of channel statistics such as Doppler/delay properties and spatial receive filtering pre-defined by a QCL type. The source RS determines the beam to use. The TCI states are configured by radio resource control (RRC) configuration. A physical downlink control channel (PDCCH) is received in a control resource set (CORESET), where a CORESET is indicated to a TCI state by MAC-CE signaling. The TCI state of a physical downlink shared channel (PDSCH) can be that of corresponding PDCCH, or a default TCI state. For physical uplink shared channel (PUSCH), the SRS resource indicator in an uplink DCI with uplink grants can be used for uplink beam indication.

Due to the fundamentally different nature of system operation relative to FR-1 bands, drastically different CSI acquisition framework is developed and optimized in 3GPP from Release 15 to 17 [4,6]. This concludes our evolution summary and leads the discussion into the massive MIMO technologies employed in Release 18 (and beyond) study item considerations.

IV. MASSIVE MIMO TECHNOLOGIES FOR 3GPP RELEASE 18 (5G-ADVANCED) AND BEYOND

The massive MIMO concepts standardized in 3GPP Release 15, 16 and 17 presents an ideal opportunity for further development in Release 18, where the massive MIMO evolution is likely to transition from a network-centric architecture to a UE-centric architecture. In this section, we discuss the key massive MIMO technologies for Release 18 and beyond across both FR-1 and FR-2 bands, including CJT for multi-TRP operation, uplink CSI enhancements, enhancements in the DMRS design, CSI for mobility enhancements, and enhancements on mmWave systems.

*A. CJT for Multi-TRP for FDD systems*

A typical scenario of CJT by multiple TRPs is illustrated in Fig. 5. There is a *coordination TRP set* (TRPs inside the black solid line circle as an example in the figure), a *CSI measurement TRP set* (TRPs within the dashed red lines inside the black circle), and *a CJT TRP set* (generally same as CSI measurement TRP set *or a subset* of CSI measurement TRP set). A fundamental problem is the determination of *optimal* coordination TRP set and CSI measurement TRP set, chosen to satisfy some optimization constraints in a UE-centric manner. It is assumed that sufficient backhaul connectivity is in place for TRPs within the coordination TRP set, while the CSI measurement TRP set and CJT TRP set can be selected in a UE-centric fashion. The CSI measurement TRP set is configured by RRC based on the RSRP *difference* with the serving cell, such that the TRPs with *strongest* RSRP are included in the CSI measurement set. Furthermore, each UE needs to measure the CSI of TRPs within the CSI measurement TRP set and report the measurement to the BS. Subsequently, the BS can determine the CJT TRP set for each UE depending on system scheduler and CSI.

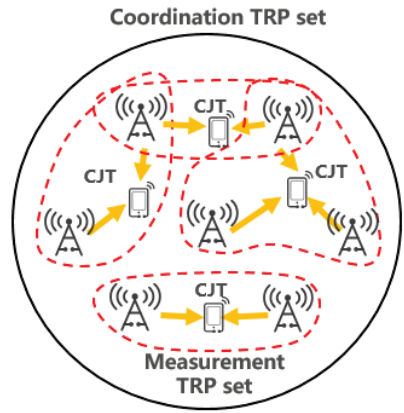

**Fig. 5.** Illustration of CJT for Multi-TRP-based systems.

The above system can function in several practical deployment scenarios tailored for CJT operation: In the first case, three RRHs are distributed and connected to the BS within a cell as shown in Fig. 6. Compared to the *single-TRP* layout, the additional three RRHs in each cell are located at the midpoints between the centers of its associated site and neighboring sites, respectively, and the antenna angle of each TRP is facing towards the center region of the associated cell. In the second scenario, inter-cell CJT exists, where multiple cells form a CJT transmission set can be considered. Figure 6 shows two representative scenarios of intra-site inter-cell and inter-site CJT inter-cell scenarios with three-cell CJT transmission sets.

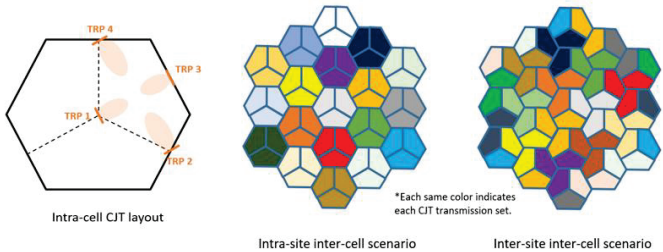

Fig. 6. Illustration of **i**ntra-site inter-cell and inter-site inter-cell CJT scenarios.

In such scenarios, to select an optimal TRP coordination set and CSI measurement TRP set, while attempting to achieve precise coherent transmission and interference suppression between multiple TRPs, *precision* and *accuracy* of channel information is critical for each UE. Thus, a high-resolution CSI feedback link needs to be designed for CJT in Release 18 for higher performance [25]. On the other hand, channel measurement and reporting of higher number of TRPs leads to increasing of both downlink RS resource and uplink feedback overhead. Therefore, the CSI enhancement needs to achieve fine CSI precision with *limited* downlink and uplink overhead. This is a significant challenge when coupled with real-time implementation constraints. The work in [25] proposes joint or separate codebook design across multiple TRP for CJT. The idea is to minimize feedback overhead allocation across multiple TRPs. To optimize feedback overhead allocation, artificial intelligence (AI)-based feedback reduction technique for FDD-based cell-free systems is proposed in [25]. Below, we present an alternative to design CJT codebook which assists in obtaining optimal TRP coordination set and CSI measurement TRP set, respectively.

The starting point of CJT codebook design can be represented as the following formulation (5), assuming a system with $N$ TRPs. Note that $P$ denotes the number of antenna ports for each TRP, and $N_f$ denotes the number of frequency-domain units. Furthermore, $L_n$ and $M_n$ are the number of spatial basis for each polarization and the number of frequency basis corresponding to the $n$-th TRP. Since the angle-delay properties are different for channels for different TRPs, the channel projection to spatial domain and the frequency domain basis is performed separately for different TRPs. The sub-matrix $W_n\ for\ n=1,2,\ldots,N$ has dimensions $P \times N_f$ corresponding to the $n$-th TRP in space-frequency form, and *is represented and approximated by the combination of a set of spatial basis $W_{n,1}\ for\ n=1,2,\ldots,N$ and a set of frequency basis $W_{n,f}\ for\ n=1,2,\ldots,N$*. Note that $W$ here is the $NP \times N_f$ joint feedback matrix in space-frequency form, following the structure given in (1). For the $n$-th TRP, the matrix $W_1$ in (1) is represented by $W_{n,1}$ is a $P \times 2L_n$ matrix consisting of $2L_n$ DFT based spatial basis, $W_{n,f}$ is an $N_f \times M_n$ matrix composed of $M_n$ DFT based frequency basis, and the matrix $W_2$ in (1) represented by $W_{n,2}$ is the space-frequency combination complex coefficients with dimension $2L_n \times M_n$. For each TRP, $K_n$ strongest coefficients are selected from all $2L_nM_n$ coefficients for reporting. Note that the values of $L_n$, $M_n$ and $K_n$ can be the same or different for each TRP. A special case of the above codebook design is to utilize joint frequency-domain compression across TRPs, where frequency basis vectors are commonly selected for all TRPs. One realization of the codebook design can then be formulated as

$$W = \begin{bmatrix} W_{1,1}W_{1,2} \\ \vdots \\ W_{N,1}W_{N,2} \end{bmatrix} W_f^H, \qquad (5)$$

where $W_f$ is a $N_f \times M$ matrix composed of $M$ common frequency-basis vectors for all TRPs with $M = M_n\ for\ n = 1,2,\ldots,N$. The underlying principle of the joint frequency-domain compression codebook design is to exploit the fact that channels for all TRPs can be correlated in the frequency domain, since the same bandwidth associated with configured sub-bands (SBs) are used for all the TRPs.

For further improving the performance of coherent transmission, full channel information compared with conventional eigenvector-based PMI feedback can be fed back to BS to facilitate greater interference suppression.

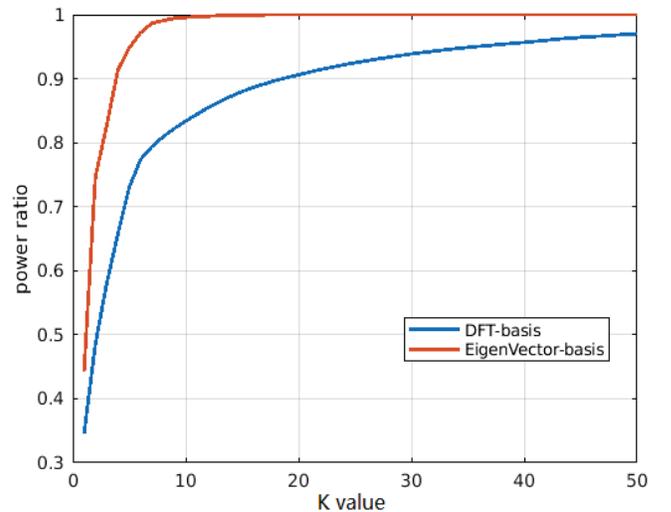

Fig. 7. The power ratio with different $K$ for different basis design

For the traditional codebook design, the spatial domain basis ($W_{n,1}$ in (5)) is a set of oversampled and orthogonal 2D-DFT beams, and the frequency domain basis ($W_{n,f}$ in (5)) is a set of orthogonal DFT vectors. *Channel feedback based on spatial and frequency-domain basis is to utilize the correlation of channel in a specific domain and to seek a set of sparse basis for an approximate representation of the beamforming matrix.* The sparser the projection of the channel on the basis, the fewer the number of basis required to represent the channel. For CJT, the *eigenvectors* of spatial (*s*) or frequency (*f*) domain statistical covariance matrix, $\mathbb{E}(\boldsymbol{H}(s,f)\boldsymbol{H}(s,f)^H)$ is much sparser than 2D-DFT basis. To compare the sparsity of different basis design, a power ratio, *r*, is introduced and can be calculated by following steps.

- ✦ **Step 1.** Calculate all the candidate coefficients with

$$\boldsymbol{C} = \boldsymbol{W}_1^H \boldsymbol{H}(s,f)\boldsymbol{W}_f, \qquad (6)$$

where $\boldsymbol{H}(s,f)$ is the union channel matrix in space-frequency form with dimension of $P \times N_f$ and $\boldsymbol{C}$ denotes the coefficient matrix by projection to the spatial and frequency domain with the dimension of $P \times N_f$. $\boldsymbol{W}_1$ ($P \times P$) and $\boldsymbol{W}_f$ ($N_f \times N_f$) are spatial and frequency basis respectively.

- ✦ **Step 2.** Choose *K* coefficients with the largest amplitude across all the coefficients in $\boldsymbol{C}$ matrix which can be donated as $C_k$ from $k = 1,2,…,K$.
- ✦ **Step 3.** Compute the power ratio of the sum of powers of the *K* coefficients relative to the sum power of all coefficients in $\boldsymbol{C}$, i.e., $\boldsymbol{C}(i,j)$ for $i = 1,2,…,P$ and $j = 1,2,…,N_f$. Note that here $\boldsymbol{C}(i,j)$ denotes the $(i,j)$-th element of $\boldsymbol{C}$. The power ratio can be expressed as

$$r = \frac{\sum_{k=1 \sim K} \|C_k\|^2}{\sum_i \sum_j \|\boldsymbol{C}(i,j)\|^2}. \qquad (7)$$

The expression in (7) correlates to accuracy of reconstructed CSI through CSI feedback at BS. An important observation is that the larger power ratio, the higher the accuracy of recovered CSI at BS. The power ratios for different value of *K* with different basis (DFT basis and eigenvector basis) are calculated and shown in Fig. 7. One can readily observe from the results that compared with DFT basis, a greatly reduced number of coefficients are needed to achieve same power ratio, since the eigen-basis is able to accurately match the UE-specific statistical subspace better than 2D DFT basis. Even for extreme cases like $K > 50$, the power ratio remains approximately 0.95. Therefore, if the spatial and/or frequency domain basis in the codebook structure in (5) are selected from the *statistical eigenvectors* for the corresponding domain, the CSI precision can be optimized.

### B. Coherent JT for MTRP for TDD systems

While CJT brings attractive merits, and it also poses higher requirements on the accuracy of CSI obtained on SRS, since the beamformers of the coordinated cells are required to assure that the phase of received signals from different coordinated cells should be constructive based on the obtained channel information. One important factor that limits the channel estimation accuracy on SRS is the *inter-cell interference*. The current interference suppression method is to whiten the colored interference by low correlation sequences during LS operation at BS. However, the sequence correlation is limited by sequence length and can hardly to be improved. For example, the cross correlation between different root sequences is larger if the length of the root sequence is shorter. To achieve better SNR for SRS, frequency hopping would be performed, and the shorter root sequence would be used. In that case, poor LS performance would be seen.

The SRS signal transmitted by a UE would be received and estimated by all of its coordinated TRPs to enable CJT transmission. The distributed TRPs will cause uneven RSRP of received SRS signals. As seen in Figure 8a), the SRS of UE1 to its coordination cell suffers severe interference from a near located UE2. The performance decrease caused by poor estimation of SRS signal in multi-TRP case, which can be seen from the 4 dB gap of mean squared error (MSE) in Figure 8b) compared with that of single-TRP case. The MSE for each SRS channel estimation used in this section is defined as:

$$\text{MSE}_p = \frac{\sum_{k,s} |\hat{H}_{p,k,s} - H_{p,k,s}|^2}{\sum_{k,s} |H_{p,k,s}|^2}, \qquad (8)$$

where $\hat{H}_{p,k,s}$ and $H_{p,k,s}$ are the estimated channel coefficient and ideal channel coefficient corresponding to SRS port *p*, TRP receiving antenna *k* and subcarrier *s*.

In the simulation, 4 UEs per cell is assumed and totally 21 cells are setup. One 4-port SRS (for sounding 4 receive radio chains) is transmitted by each UE for all of its coordinated cells. Each UE has one serving cell and multiple coordinated cells, the coordinated cells are determined to satisfy that the RSRP gap between the serving cell and the coordinated cell should be no more than 10 dB. The physical resource of the SRS is allocated by each serving cell. For example, orthogonal physical resources and same root sequence are used for each UE in the same serving cell. All cells share the same set of SRS resources, and different cells would use different root sequences. The blue line in Fig. 8b) collects the MSE from all of the serving cells via SRS channel estimation, and the red line collects the MSE from all of the coordinated cells. The yellow line assumes that no interference but only the noise is suffered by the coordinated cells. To resolve the inter-cell SRS interference, especially for the case with limited SRS resource in the system, SRS interference randomization should be introduced to improve the SRS channel estimation performance under the scenario with strong interference.

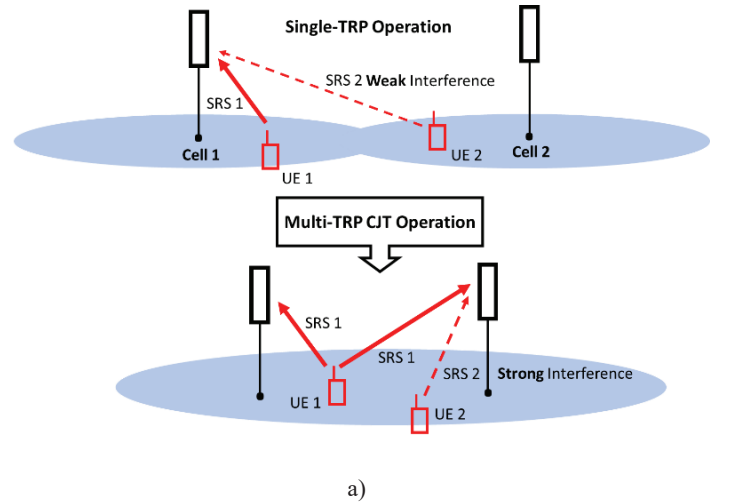

a)

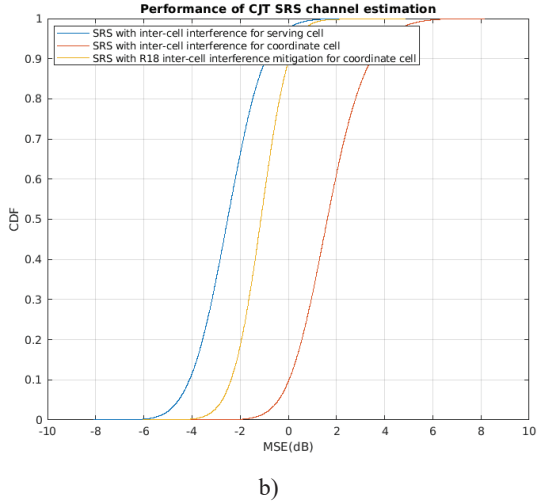

**Fig. 8.** a) Multi-TRP scenario due to inter-cell interference; b) MSE performance (defined as per (8)) for the type of scenario in a).

Assuming that one target UE and one strong interference UE transmit SRS to a TRP on the same physical resource, the received SRS signal model at the $n$-th transmission in the point of view of the TRP can be written as

$$y_n(m) = r_n(m)h_n(m) + r'_n(m)h'_n(m) + w_n(m)$$
$$= e^{j\alpha_n m}\bar{r}(m)h_n(m) + e^{j\alpha'_n m}\bar{r}'(m)h'_n(m)$$
$$+ w_n(m). \quad (9)$$

where $(m)$ denotes the $m$-th element of a sequence mapping to different subcarriers in frequency domain, $m = 0,1,\ldots,M\text{-}1$, $M$ is the number of subcarriers occupied by the SRS. $y_n$ and $w_n$ are the received SRS sequence and the noise. $r_n$ and $r'_n$ are the SRS sequence of the target UE and that of the interference UE. $h_n$ and $h'_n$ are the target channel and the interference channel coefficient in frequency domain. $\bar{r}$ and $\bar{r}'$ are the sequence used by target UE and the interference UE. $\alpha_n$ and $\alpha'_n$ are the correlation CS values for the two UEs, which are randomized among multiple transmissions. The CS value can be set between 0 to $2\pi$. To estimate the target channel, $y_n$ is multiplied by the conjugate of $r_n$, then we have

$$\tilde{y}_n(m) = y_n(m)r_n^*(m)$$
$$= h_n(m) + r_n^*(m)r'_n(m)h'_n(m) + r_n^*(m)w_n(m), \quad (10)$$

where $r_n^*(m)r'_n(m)h'_n(m) = e^{j(\alpha'_n - \alpha_n)m}\bar{r}^*(m)\bar{r}'(m)h'_n(m)$. Then we transform the above signal to the delay domain, giving:

$$\tilde{y}_n^D(\tau) = IFFT(\tilde{y}_n) = h_n^D(\tau) + I_n^D(\tau) + w_n^D(\tau), \quad (11)$$

where $h_n^D(\tau), I_n^D(\tau)$, and $w_n^D(\tau)$ are the target channel, interference, and noise in the delay domain, respectively. Furthermore, we denote $\bar{I}_n^D(\tau) = IFFT(\bar{r}^* \circ \bar{r}' \circ h'_n)$ to be the delay-domain interference without adding and removing CS, which is not whitened. Since a frequency domain cyclic phase shift is a delay shift in the delay domain,

$$I_n^D(\tau) = \bar{I}_n^D\left(\tau + \frac{\tau_{max}}{\alpha_{max}}(\alpha'_n - \alpha_n)\right), \quad (12)$$

where $\tau_{max}$ is the maximum delay within an OFDM symbol, $\alpha_{max}$ is the maximum CS value. By time filtering across multiple SRS transmissions (here we take time-average as an example), we calculate the statistical power-delay profile (PDP) of the received signal:

$$\text{PDP}_{\tilde{y}}(\tau) = \frac{1}{N}\sum_{n=1}^{N}|\tilde{y}_n^D(\tau)|^2. \quad (13)$$

For a large $N$, assuming that the target signal and the interference are independent,

$$\text{PDP}_{\tilde{y}}(\tau) = \frac{1}{N}\sum_{n=1}^{N}|h_n^D(\tau)|^2 + \frac{1}{N}\sum_{n=1}^{N}|I_n^D(\tau)|^2 + \sigma_n^2 + o(1)$$
$$= \text{PDP}_h(\tau) + \text{PDP}_I(\tau) + \sigma_n^2 + O(1), \quad (14)$$

where $\sigma_n^2$ is the noise variance, $O(1) \to 0$ when $N \to \infty$. Due to the fact that the channel consists of only a few paths and the path delays vary slowly, the power of $\text{PDP}_h(\tau)$ concentrates on only a few strong delay taps. If the power of $\text{PDP}_I(\tau)$ spreads over the entire delay domain, the interference and noise and be reduced by finding the strong delay taps of $\text{PDP}_{\tilde{y}}(\tau)$. However, $\text{PDP}_I(\tau)$ may also have strong delay taps in practice, which causes severe channel estimation error. To solve the problem, CS hopping is needed to whiten $\text{PDP}_I(\tau)$.

Without CS hopping, $\alpha'_n - \alpha_n$ is constant among different SRS transmissions. Then

$$\text{PDP}_I(\tau) = \frac{1}{N}\sum_{n=1}^{N}|I_n^D(\tau)|^2 = \frac{1}{N}\sum_{n=1}^{N}|\bar{I}_n^D(\tau + b)|^2, \quad (15)$$

where $b$ is a constant. $\text{PDP}_I(\tau)$ varies with $\tau$, so it is not white in the delay domain.

However, by adopting CS hopping, $\text{PDP}_I(\tau)$ can be whitened:

$$\text{PDP}_I(\tau) = \frac{1}{N}\sum_{n=1}^{N}\left|\bar{I}_n^D\left(\tau + \frac{\tau_{max}}{\alpha_{max}}((\alpha'_n - \alpha_n)\backslash\alpha_{max})\right)\right|^2. \quad (16)$$

Since $(\alpha'_n - \alpha_n)\backslash\alpha_{max}$ is uniformly distributed in $[0, \alpha_{max})$, if infinity transmission times is assumed, each tap in PDP has equal probability to be occupied, so we can get:

$$\lim_{N\to\infty}\text{PDP}_I(\tau) = \frac{1}{\alpha_{max}}\int_0^{\alpha_{max}}\left|\bar{I}_n^D\left(\tau + \frac{\tau_{max}}{\alpha_{max}}\alpha\right)\right|^2 d\alpha$$
$$= \frac{1}{\tau_{max}}\int_0^{\tau_{max}}|\bar{I}_n^D(\tau')|^2 d\tau'. \quad (17)$$

We can see that $\text{PDP}_I(\tau)$ is independent to $\tau$, which means that $\text{PDP}_I(\tau)$ is whitened by adopting CS hopping. Therefore, the strong taps of the target signal can be found correctly with the time filtering. For example, the taps with larger power than the power of whitened interference taps added by noise power can be easily found. The performance of proposed CS hopping is presented in Fig. 9. In Figure 9a), we collect the error values of estimated tap compared with the ideal tap for each channel estimation. Both the CDF of estimation error values of the strongest tap and the estimation error values of the first three strongest taps are presented. It can be observed that via CS hopping, the taps of target signal can be found more accurately due to that the interference is whitened. According to the MSE of channel estimation in Fig. 9b), we can further see that channel estimation can be performed more accurately compared with the legacy case.

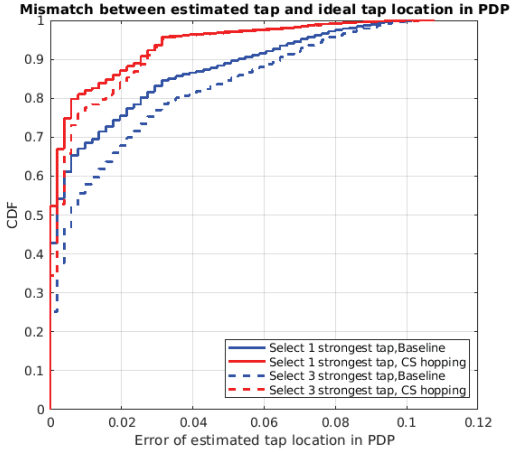

a) Comparison between the estimation error of tap location for the case w/. and w/o. CS hopping.

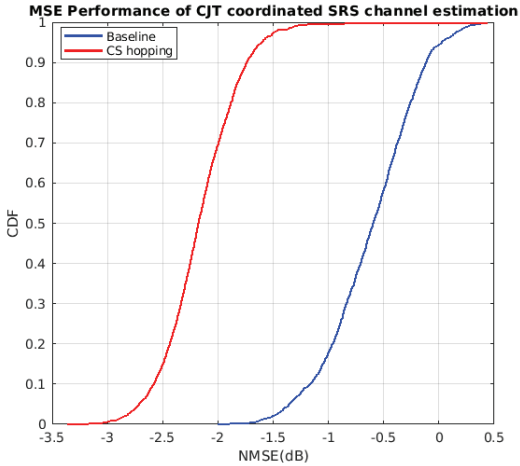

b) MSE of SRS channel estimation for the case w/. and w/o. CS hopping

**Fig. 9.** Illustration of the performance benefits of CS hopping for interference randomization

In TDD, thanks to the channel reciprocity, the uplink channel estimation can be used to infer downlink channels. This feature enables BS to reduce the feedback overhead. However, due to RF mismatch in the uplink and downlink paths (at receivers and transmitters), utilizing the uplink channels for downlink channels requires periodic *calibration* among receive and transmit antennas of RF networks at the BS. In general, a BS has an on-board calibration mechanism in its own RF network to calibrate its antenna panels with multiple receiver and transmitter antennas. The on-board calibration mechanism can be performed via small-power reference signal transmission and reception from/to the RF antenna network of base station in order to measure the gain and phase differences among transceivers in the same RF unit. Thus, it can be done by hardware implementation in a self-contained manner. However, it can be difficult to perform such an on-board calibration with non-collocated TRPs. Instead of on-board calibration, an *over-the-air signaling mechanism* to calibrate receive and transmit antennas among non-co-located TRPs is instrumental. In this regard, an RS-based mechanism for a UE-assisted calibration can be used to report calibration coefficients *across* TRPs.

### C. Enhancements on Uplink MIMO

The accuracy of uplink CSI/beamforming and the allowed maximum transmission layers are *essential* for the uplink spectrum efficiency. With the emergence of advanced UEs types such as hubs for FWA, router-like devices for customer premise equipment (CPE), industrial robotics, and automobiles, larger form factors become feasible for UE antenna architectures. This not only allows larger number of antenna elements, but also facilitates increased number of transmission layers, translating to higher peak data rates and improved uplink coverage.

*1) CSI Enhancements on the Uplink*: Similar to CSI for downlink, uplink precoding signaling design depends on whether downlink-uplink reciprocity is available or not. When reciprocity is absent, the BS determines the uplink precoding for a UE by measuring uplink reference signals such as SRS. For NR, this mechanism, known as the codebook-based uplink transmission, includes the signaling of uplink precoding information via downlink control signaling. Only coarse codewords consist of 1/-1/j/-j are employed in codebook-based uplink transmission due to the indication overhead limitation via downlink control signaling. When reciprocity is present, a UE can measure downlink reference signals (such as CSI-RS) to determine uplink precoding. For NR, this mechanism, known as the non-codebook-based uplink transmission, employs the signaling of SRS port selection via downlink control signaling, where the uplink precoder weighted on uplink SRS is chosen as uplink precoder for uplink data transmission. However, the candidate uplink precoding is calculated by UE without jointly considering multi-user interference. The trade-off between downlink indication overhead and uplink precoding accuracy is always an issue during uplink codebook design. An efficient downlink indication method is proposed in this section by indicating uplink precoder on downlink CSI-RS. Compared with codebook-based uplink transmission, high resolution uplink precoding can be indicated with low indication overhead. Compared with codebook-based uplink transmission, the uplink precoding is determined at BS side and multi-UE interference can jointly considered. Specifically, with the proposed scheme, the precoder for each uplink data stream is indicated by a weighted CSI-RS transmitted by BS, where weighted CSI-RS means CSI-RS is transmitted by BS with a weight or precoder.

Assuming TRP-to-UE link channel is $\boldsymbol{H}$ with the dimension of $n_{Rx} \times n_{Tx}$, where $n_{Rx}$ and $n_{Tx}$ are the size of receive antenna at UE and transmit antenna at BS, the received signal for the weighted CSI-RS can be expressed as:

$$\boldsymbol{y} = \boldsymbol{H}\boldsymbol{W}_{DL}, \qquad (18)$$

where $\boldsymbol{y}$ is the received signal for the weighted CSI-RS with dimension of $n_{Rx} \times 1$ and $\boldsymbol{W}_{DL}$ with the dimension of $n_{Tx} \times 1$ is the weight or precoder for CSI-RS. Considering the dimension of the received signal **y** being $n_{Rx} \times 1$, the received signal for the weighted CSI-RS can be used as a UL precoder for UE if a appropriated weight or precoder for CSI-RS $\boldsymbol{W}_{DL}$ is designed. Then,

$$\boldsymbol{P}_{UL} = \boldsymbol{y} = \boldsymbol{H}\boldsymbol{W}_{DL}. \quad (19)$$

High resolution $\boldsymbol{P}_{UL}$ is calculated at BS side and can be indicated to UE by choosing an appropriated $\boldsymbol{W}_{DL}$ and the received signal **y** can be interplated as UL prcoder. Innumerable solutions for $W_{DL}$ can be found in (19) due to the number of transmit antennas being greater than the number of receive antennas, and the one with largest $\|\boldsymbol{H}\boldsymbol{W}_{DL}\|^2$ can be chosen as weight on CSI-RS to maximize the CSI-RS coverage. Then the DCI indication overhead can be reduced, and the resolution of indication can be increased. Besides, the uplink precoder for each UE is calculated at BS side and can be jointly optimized to maximize uplink throughput considering multi-UE interference.

*2) DMRS Enhancement for the Uplink*: In 5G-NR, DMRS is used for channel estimation and demodulation of physical channels [13]. Two types of DMRS are standardized in 3GPP for NR use till Release 17, namely Type-I DMRS and Type-II DMRS, as shown in Fig. 10. Type-I DMRS is more denser in the frequency domain relative to Type-II, and both types support up to 8 to 12 orthogonal DMRS ports. However, in 3GPP Release 18 (5G-Advanced) and future releases, more receive antennas would be employed at each BS and joint multi-TRP reception would be enabled with more advanced network deployments. To this end, more orthogonal DMRS ports are needed for more uplink transmission layers. Increasing the DMRS ports without increasing DMRS transmission overhead is a major challenge while considering DMRS design.

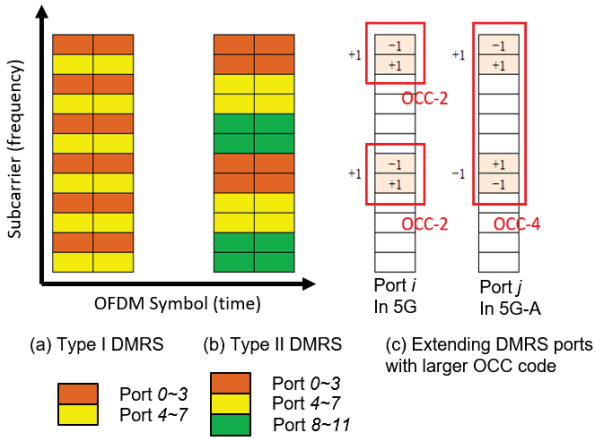

**Fig 10.** DMRS patterns in 5G until 3GPP Release 17 and the proposed DMRS enhancements for 3GPP Release 18.

To increase the number of DMRS ports, multiplexing DMRS in the time delay domain is a promising solution. The data channel for each UE may have considerable limited delay spread variability, and DMRS from multiple UEs can be transmitted in the same time-frequency resource by moving the signal in the time delay domain, as shown in Fig.11. The delay spread of each UE is limited and can be shifted in the time delay domain without changing channel property. The DMRS ports from different UE are transmitted in the same time-frequency resource, while can be distinguished in the time delay domain if the delay spread of multiple UEs are not overlapped. Shift in the time delay domain is equivalent to multiplexing DFT code in the frequency domain. To extend the number of orthogonal DMRS ports from 12 to 24, a simple scheme is doubling the length of orthogonal cover code (OCC) in 5G from 2 to 4.

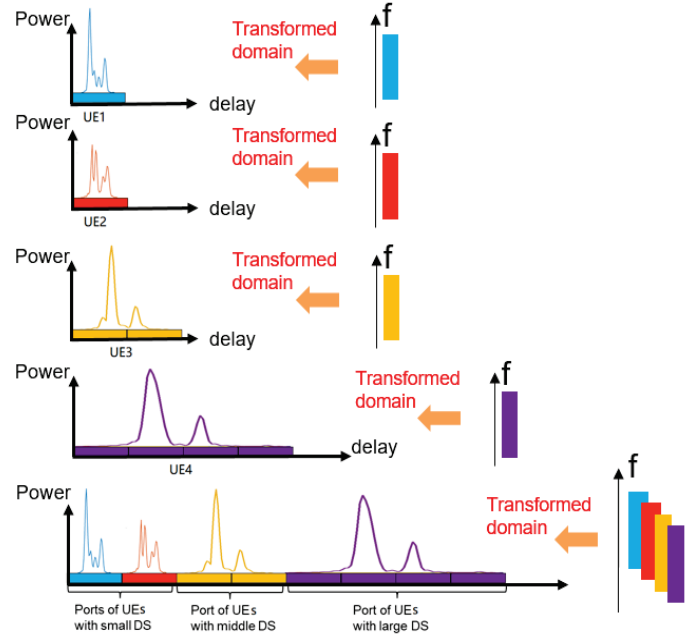

**Fig. 11.** Multiplexing DMRS ports for different UEs in time delay domain.

### D. CSI Enhancements for Mobility

In existing CSI measurement and feedback procedures till 3GPP Release 17, the BS generally assumes that the channel condition remains unchanged within the CSI reporting period, and the BS uses the latest reported channel CSI for downlink transmission beamforming. However, the performance in scenarios involving mobility is greatly degraded due to more rapid CSI expiration/outdatedness. Here, the channel may vary rapidly, and the CSI expiration significantly deteriorates the system performance. In addition, this becomes worse in multi-UE scenarios, as the CSI mismatches will introduce the inter-UE interference when multi-UE pairing is performed by the system scheduler. The expiration process based on CSI-RS and PMI feedback is illustrated in Fig. 12.

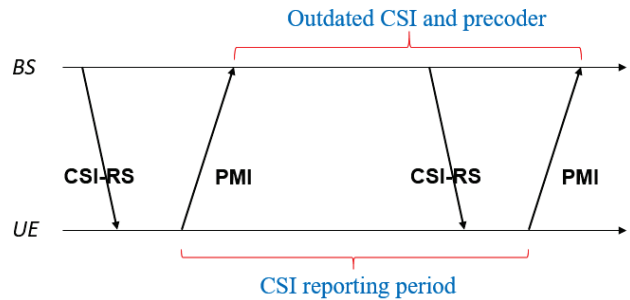

**Fig. 12.** Conventional CSI reporting in 3GPP Release 15 to 17.

This above motivates new solutions for CSI enhancements with mobility in 3GPP Release 18. In massive MIMO transmission, a channel consists of multiple paths, and the time-varying characteristic of a single path can be uniquely described by the exponential function proportional to its Doppler frequency:

$$H(s,f,t) = \sum_{i=1}^{L} \alpha_i e^{j2\pi v_i t} \boldsymbol{\theta}_i \otimes \boldsymbol{\tau}_i, \quad (20)$$

where $H(s,f,t)$ means the union channel matrix in space-frequency form with dimension of $P \times N_f$ for slot $t$. Note that $\alpha_i e^{j2\pi v_i t}$ denotes the channel coefficient for $i$-th path, where $\alpha_i$ is irrelevant to mobility and $v_i$ denotes the equivalent Doppler shift for $i$-th path. Furthermore, $\boldsymbol{\theta}_i$ ($P \times 1$) and $\boldsymbol{\tau}_i$ ($1 \times N_f$) are angle and delay steering vectors for $i$-th path, respectively. The time-frequency channel matrix, $H(s,f,t)$, at slot $t$ can be transformed into angel-delay domain by 2-demsion IDFT, as each yellow rectangular in figure X.

$$C = W_1^H H(s,f,t) W_f, \quad (21)$$

where $C$ is angle-delay domain complex coefficient matrix with each coefficient (each red box in figure X) corresponds to one angle-delay pair. $W_1$ (P × P) and $W_f$ ($N_f \times N_f$) are spatial and frequency basis respectively. The Doppler information can be extracted from the complex coefficients of continuous slots for each angel-delay pair, as shown in Figure 13.

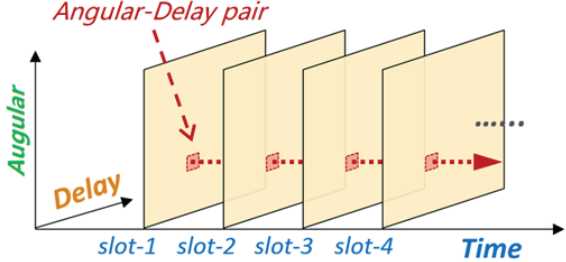

**Fig. 13.** CSI enhancement concepts for mobility in 3GPP Release 18.

As the Doppler frequency does not change rapidly in short time duration, CSI reporting in the mobility scenarios can be enhanced by predicting the time-varying channel based on the Doppler frequency of each path. The procedure for CSI enhancement in mobility case is shown in Fig. 14 and can be summarized as 3 steps.

- **Step 1.** *BS sends continuous N CSI-RS to UE:*
  - The time gap between two adjacent CSI-RS transmissions can be denoted by $\Delta t$, which is determined by Doppler spread. The whole CSI-RS occupies $N \cdot \Delta t$, which is determined by the required Doppler resolution.
- **Step 2.** *Doppler extraction and CSI prediction:*
  - Step 2.1: $\alpha_i$ and the Doppler information $v_i$ (i=1~L) can be extracted by $N$ union channel matrix in space-frequency form (see the formulation above (20)) estimated from $N$ CSI-RS in Step 1.
  - Step 2.2: UE can predict the CSI information in future M slots based on $e^{j2\pi v_i t}$ and consistent $\alpha_i$.
- **Step 3.** UE do CSI compression on the predicted CSI via Doppler domain.

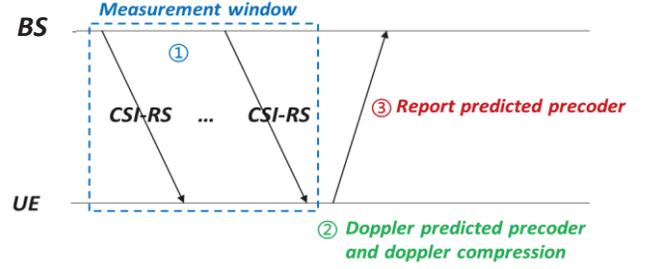

**Fig. 14.** 3GPP Release 18 proposed procedure for CSI measurement and reporting for mobility scenarios based on CSI-RS reference signals.

In Step 3 above, to compress the predicated CSI ($P \times N_f$) in future $N_{slot}$ slots, the codebook for CSI reporting can be expressed as follows considering CSI compression across spatial domain, frequency domain, and time domain:

$$W = W_1 W_2 (W_f \otimes W_D)^H, \quad (22)$$

where $W$ means the predicted CSI or precoder in future $N_{slot}$ slots with dimension of $P \times (N_f \cdot N_{slot})$, $W_1$, $W_f$ and $W_D$ means the spatial domain basis, frequency domain basis and time domain basis for CSI compression. $W_1$ and $W_f$ reuse Release 16 or Release 17 design. $W_1$ is a $P \times 2L$ matrix consisting of 2L DFT based spatial basis, $W_f$ is an $N_f \times M$ matrix composed of M DFT based frequency basis. $W_D$ is an $N_{slot} \times T$ matrix composed of $T$ DFT − based time basis and $W_2$ is the space- frequency-time combination complex coefficients with dimension $2L \times (M \cdot T)$. K strongest coefficients are selected from all $2L \times (M \cdot T)$ coefficients for reporting.

### E. Enhancements on mmWave/FR-2 Bands

In 3GPP Release 15 and 16, a common framework is shared for CSI acquisition (FR-1) and beam management (FR-2). While the complexity of such a framework is justified for CSI in FR1, it makes beam management procedures less efficient in FR2. Efficiency here refers to the overhead associated with beam management operations and latency for reporting and indicating new beams, which in turn impact reliability. Furthermore, as aforementioned, the beam management procedures can be different for different channels. Having different beam indication/update mechanisms increases the complexity, overhead, and latency of beam management. Such drawbacks are especially troublesome for high mobility scenarios (such as highway/vehicular use cases at FR-2) and/or scenarios requiring large number of configured TCI states. In these challenging scenarios, inefficiencies mentioned above would lead to not only loss of throughput, but also loss of connections. These drawbacks motivated a streamlined beam management framework for multi-beam operations and procedures that is common for data and control, and UL and DL channels. This framework, is referred to as the unified TCI framework, first introduced in Release 17 for single TRP operation and is being enhanced in Release 18 for multi-panel TRP operation.

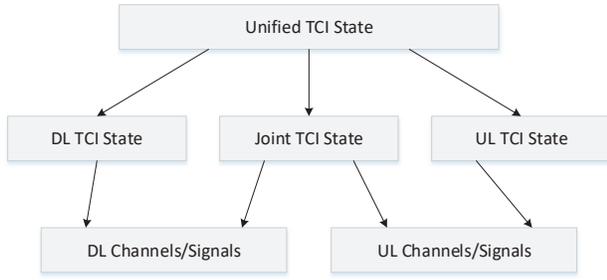

**Fig. 15.** Illustration of association of unified TCI state with downlink (DL) and uplink (UL) channels and signals.

The unified TCI framework supports signaling a unified TCI state to the UE, where a unified TCI state is, as illustrated in Fig Y1:
- A downlink or a Joint TCI state, where a downlink TCI state is applied to downlink channels and signals, and a joint TCI state is applied to downlink channels and signals, , and/or
- An uplink or a Joint TCI state, where an uplink TCI state is applied to UL channels and signals.

The unified TCI state is applied at least UE dedicated-channels, which are channels that are transmitted to a single-UE or channels transmitted from a UE. The unified TCI state framework is designed to support downlink receptions and uplink transmissions in the UE with a joint (common) beam indication for downlink and uplink by leveraging beam correspondence as illustrated in Fig. 16 (a) as well as downlink receptions and uplink transmissions in a the UE with separate beam indications for downlink and uplink, as illustrated in Fig. 16 (b), for example to mitigate maximum permissible exposure, where the beam direction of an uplink transmission is different from the beam direction of a downlink reception to avoid exposure of the human body to radiation. The unified TCI state can be one of:

1. In case of joint TCI state indication, wherein a same beam is used for downlink and uplink channels, a joint TCI state that can be used for downlink channels and uplink channels.
2. In case of separate TCI state indication, wherein different beams are used for downlink and uplink channels, a *downlink* TCI state that can be used for uplink channels.
3. In case of separate TCI state indication, wherein different beams are used for downlink and uplink channels, an *uplink* TCI state that can be used for uplink channels.

The network configures the UE two lists of a TCI states, as illustrated in Fig Y3.:
- The first list of TCI states is used for downlink and joint beam indication. Each TCI state includes a TCI state ID; two QCL information elements each including a QCL Type and a source reference signal; associated uplink power control information and pathloss reference signal.
- The second list of TCI states is used for uplink beam indication. Each TCI state includes a TCI state ID; uplink spatial relation reference signal; uplink power control information and pathloss reference signal. The uplink spatial relation reference signal can be a downlink reference signal (e.g., SSB or CSI-RS) or an uplink reference signal (e.g., SRS).

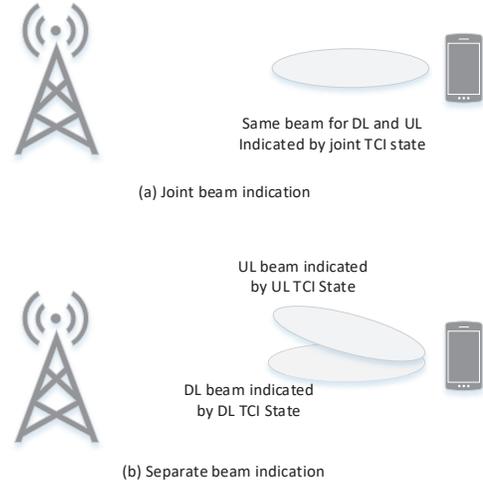

**Fig 16.** Examples of joint and separate beam indication.

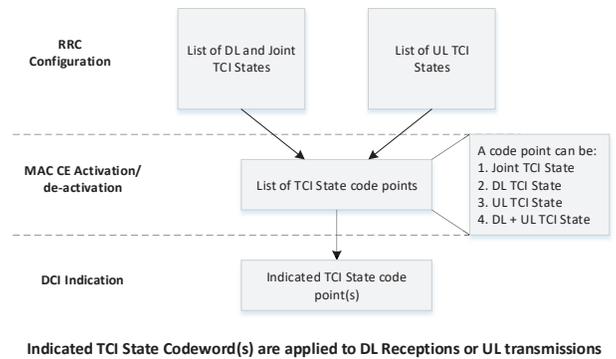

**Fig 17.** Configuration and indication of TCI states.

The network activates a set of up to 8 TCI state code points by MAC CE signaling as illustrated in Fig. 17. A TCI state code point can be (1) a downlink TCI state; (2) an uplink TCI state; (3) a joint TCI state; or (4) a pair of downlink TCI state and uplink TCI state. A TCI state code point is signaled to the UE in a downlink-related DCI Format (e.g., DCI Format 1_1 or DCI Format 1_2) using the "transmission configuration indication" field. The downlink-related DCI format may or may not include a DL assignment. The latter is a DCI Format used specifically for beam indication. As illustrated in Fig. 18, the TCI state signaled by the MAC CE is applied after a beam application time from the end of UL channel conveying HARQ-ACK feedback associated with the DCI format conveying the TCI state. If the MAC CE activates a single TCI state code point, the activated TCI state code point is applied after the MAC CE processing latency without further DCI signaling.

The unified TCI framework applies to intra-cell beam management, wherein, the TCI states have a source RS (i.e., defining a beam) that is associated with an SSB of a serving cell. The unified TCI state framework also applies to inter-cell beam management, where a TCI state can have a source RS (i.e., defining a beam) that is associated, with an SSB of a cell that has a physical cell identity (PCI) different from the PCI of the serving cell. As illustrated in Fig. 19, in Release 17, UE-dedicated channels can be received and/or transmitted using a TCI state associated with a cell having a PCI different from the PCI of the serving cell, while common channels are received or transmitted using a TCI state associated with the serving cell. The multi-TRP transmission holds promise for improving link reliability (by adding redundant communications paths for spatial diversity), system capacity (by aggregating resources from physically non-collocated TRPs), and coverage. Operating at mmWave frequencies can exploit the full potential of multi-TRP transmission; this, however, would require highly efficient beam management procedures. In Release 17, various design aspects related to beam management including beam measurement/reporting and beam failure recover (BFR) were specified for the multi-TRP operation. The corresponding TCI framework, however, was still based on Release 15/16, which has been proven inefficient and complex especially when more than one TCI states (or TRPs) are indicated. Hence, there is a need to apply/extend the unified TCI framework to the multi-TRP transmission – being enhanced in Release 18 – to fulfill the stringent latency and overhead requirements for 5G-Advanced and beyond.

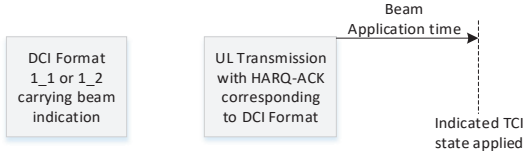

Fig. 18. Beam application time after acknowledgment of beam indication reception.

In the Release 17 unified TCI framework, a single TCI state (or a single pair of TCI states) is indicated by a TCI state codepoint of the DCI field 'Transmission Configuration Indication' in the corresponding DCI (e.g., DCI format 1_1 or 1_2 with or without DL assignment). Indicating a single TCI state/pair of TCI states can be restrictive as it is tailored for single-TRP operation. Hence, extending the Release 17 unified TCI for multi-TRP includes indication of more than one TCI states or pairs of TCI states (e.g., in a DCI format as depicted in Fig. 20), and association between TCI states and different TRPs.

## V. EVALUATION RESULTS FOR MASSIVE MIMO TECHNOLOGIES IN 3GPP RELEASE 18 AND BEYOND

In the section, we provide system-level simulation results for the key enhancements discussed in the earlier section and the evolutions discussed in previous sections. In addition, we present results from a real-world trial on multi-TRP CJT in single and multi-UE scenarios.

### A. Simulation Results for Aspects of Massive MIMO

#### a) Performance Evaluation of CJT

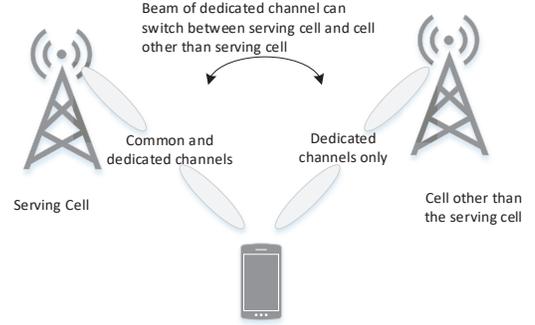

Fig 19. Support of beam switching between serving cell and cell other the serving cell.

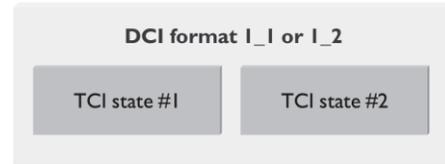

Fig. 20. A conceptual example of indicating two TCI states for two TRPs in DCI format 1_1 or 1_2.

Using the parameters given in Table A-1 in the Appendix section of the paper, figures 21 and 22 depict SLS results on the performance gains of CJT over single-TRP transmission with Release 16 Type-II enhanced codebook in intra-cell and inter-cell CJT scenarios, respectively. A total of 57 sectors (19 sites) are modelled and it is assumed that 30 UEs are randomly dropped in each sector. A non-full buffer model, FTP traffic model 1 with packet size 0.5 Mb [21], is assumed and arrival rate $\lambda$ is adjusted for a given range of resource utilization (RU) 30~40%. Each UE is assumed to be equipped with two antenna elements and up to four layers of multiuser MIMO is considered in scheduling at the BS (for either single-TRP or multi-TRP case) in each sector. Other simulation assumptions are described in Table A-1 and the simulation procedure followed is exemplified by Fig. A-1 in the Appendix section. For comparing performance of multi-TRP CJT transmission vs single-TRP transmission, user perceived throughput (UPT) is used as a performance metric. The UPT for a UE is defined as

$$r = \frac{\sum_i S_i}{\sum_i T_i}, \qquad (23)$$

where $S_i$ is the size of $i$-th data burst transmitted to the UE and $T_i$ is the time between the arrival of the first packet of the $i$-th data burst by a given data-traffic model and the reception of the last packet of the $i$-th data burst at the UE [21]. As shown in Figures A1 and A2, the CJT from multi-TRP yields significant performance improvements over single-TRP transmission in both of the scenarios, which are 37% and 57% in terms of mean UPT over all users, respectively. In addition, 66% and 81% gains of CJT are shown in terms of UPT of 5[th]-percentile user, in both of the scenarios, respectively. Both results are under low-RU regime, RU 30~40%, but it is expected that the performance improvements over single-TRP can further be increased if a high-RU regime is considered.

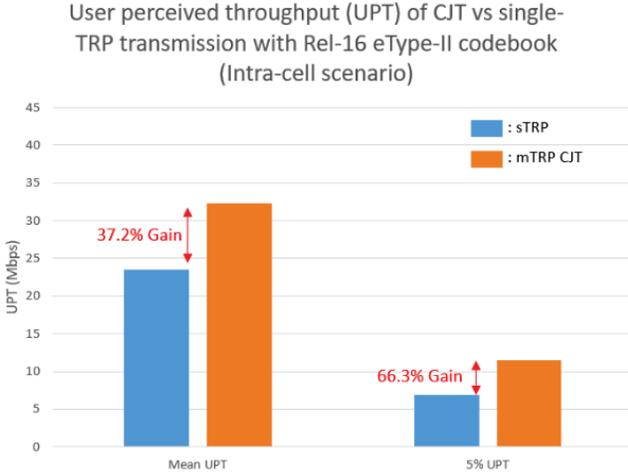

**Fig. 21.** User perceived throughput (UPT) of CJT vs single-TRP with Release16 enhanced Type-II codebook in intra-cell CJT scenario.

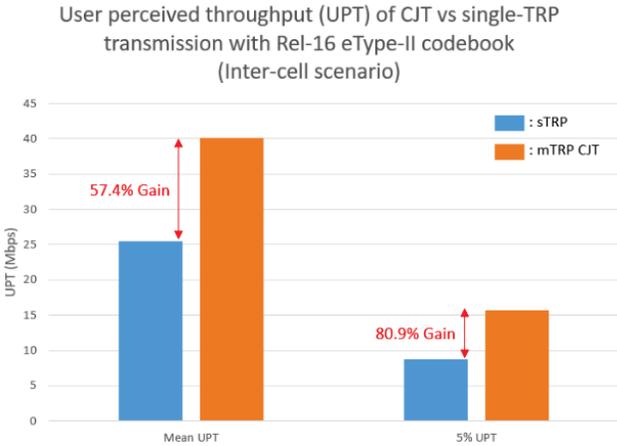

**Fig. 22.** User perceived throughput (UPT) of CJT vs single-TRP with Release 16 enhanced Type-II codebook in inter-cell CJT scenario.

In addition to the performance gain of CJT over Rel-16 codebook, SINR comparison of inter-site CJT with 3 cooperated TRPs and single-TRP transmission are provided in Fig. 23, corresponding to 32T for each TRP. Note that 4 receiving antennas are assumed at UE side. The baseline is the single-TRP transmission with Release 17 Type-II codebook. For each TRP, the overhead for the proposed CJT codebook is the same as Release 17 Type-II codebook overhead. The other detail simulation assumptions are shown in table Appendix A-I. As shown in Fig. 23, compared with single-TRP with 32 transceiver radio chains for each TRP, significant SINR improvement (about 5 dB) can be achieved by CJT with CSI enhancement. The performance gain basically comes from two aspects. One aspect is the performance gain achieved by CJT compared with no cooperation among multi-TRP, which is due to inter-cell coherent transmission and inter-cell interference suppression. Another aspect is the CJT codebook enhancement to match with the channel property of multi-TRP joint transmission.

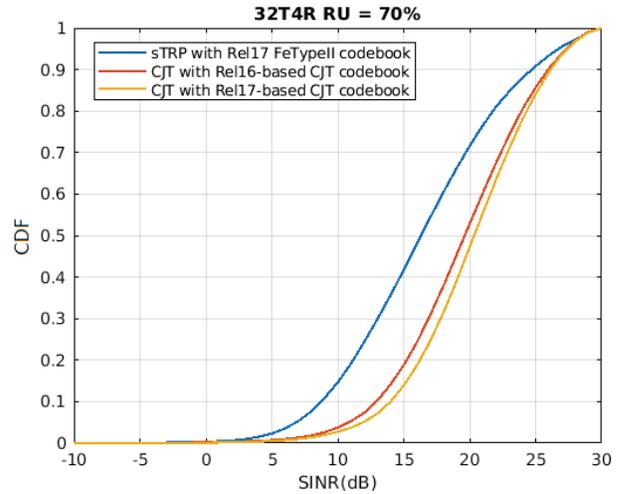

**Fig. 23.** SINR comparison of CJT for 70% RU and multi-TRP where each TRP has 32 transceiver units. Note that "s-TRP" in the figure legend refers to a single-TRP case.

b) *Performance Evaluation of Uplink MIMO Enhancement*

Figure 24 shows the uplink throughput simulation results for different UL precoding scheme under the scenario quoted in Table A-3 in the Appendix section with the respective parameters. For each TRP, 8 receive chains and for each UE, 4 or 8 transmit chains are assumed. The baseline of codebook-based uplink precoding scheme in 5G and the proposed high resolution uplink precoding with transmitting indication information on downlink CSI-RS are compared.

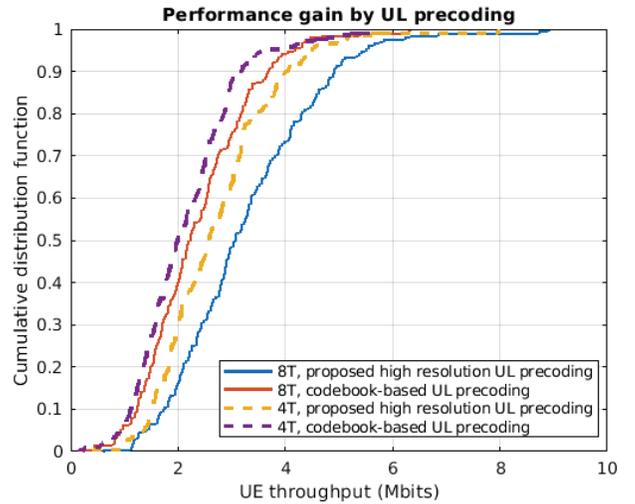

**Fig. 24.** UE throughput gain of uplink high resolution precoding with 8 and 4 transmitter chains denoted as 8T and 4T in the figure legends with UL denoting "uplink".

As shown in the result, performance gains achieved by high resolution uplink precoding can reach up to 10% for 4 transmit chains and 20% for transmit chains. With 8 transmit chains for uplink, the uplink beam will be much narrower. The performance benefit of uplink high resolution precoding becomes larger with the increase of the number of antennas ports at UE.

Figure 25 shows UL system-level simulation results for different DMRS design using the SLS conditions in Table A-3 in the Appendix, where coherent joint reception (CJR) across all TRPs is assumed. 8R for each TRP and 4T for each UE are assumed. The baseline is the type II DMRS in 5G with up to 12 orthogonal DMR ports allowed. The DMRS scheme in 5G-A employs the proposed extending OCC length scheme for DMRS enhancement. It can be observed that the throughput under 24 layers in 3GPP Release 18 are 1.75x compared with the 12 layers specification limitation in 5G. With CJR among all TRPs, UL multi-user interference can be mitigated, and more than 20 layers can be paired simultaneously.

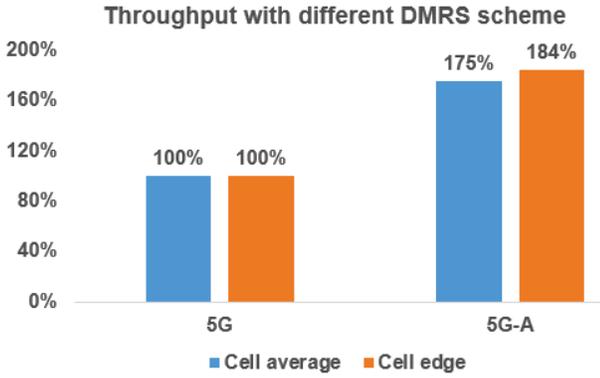

**Fig. 25.** Performance gain of high order massive MIMO in uplink.

c) *Performance Evaluation of Mobility Enhancements*

Figure 26 shows throughput simulation results for CSI predication with proposed CSI enhancement for mobility. The baseline is conventional CSI reporting with Release 16 Type-II. The new codebook structure with time domain basis is used for the proposed scheme for CSI compression on the predicted CSI/precoder in future slots. Based on simulation results, it can be observed that CSI prediction based on Release 16 type II codebook can achieve about 15% performance gain compared with Release 16 Type II conventional.

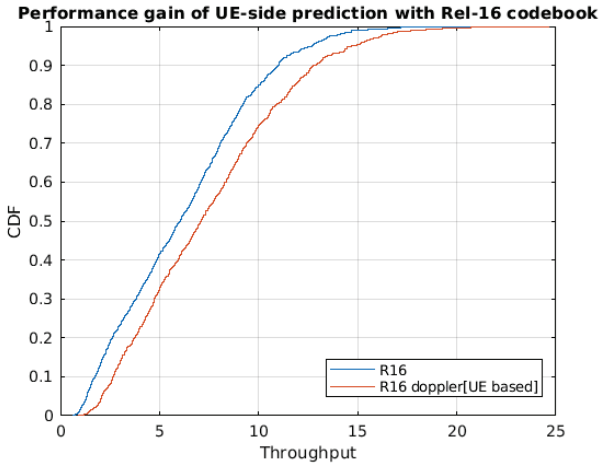

**Fig. 26.** Performance gain at the UE with Release 16 conventional and Doppler-based codebooks.

### B. *Preliminary Field Test Results*

Preliminary field test results for coherent joint transmission with multi-TRP are provided and discussed in this section, including single-user (SU) field test and multi-UE field test., where the detailed parameter is shown in Table 1. In the field testing, interference cells are also assumed and only two TRPs for coherent joint transmission are used.

Table 1 – Field Trial Specifications and Parameters.

| Parameter | Value |
|---|---|
| Location | Shanghai Jiao Tong University |
| Environment | Urban outdoor |
| Duplex Mode | TDD |
| BS Configuration and Transmit Power | 64 transmit (T) 64 receive (R), 46 dBm |
| UE Configuration | 2 transmit (T) 4 receive (R) with SRS-based antenna switching (Huawei Mate30) |
| Center Frequency | 3.7 GHz |
| Bandwidth | 20MHz |
| Transmission TRP Set | 2 TRPs with intra-site (Normal direction of the antenna as shown in yellow arrow) |
| Interfering TRP Set | 4 TRPs with inter-site and intra-site (Normal direction of the antenna as shown in red arrow) |

The SU field test scenario is shown in the Fig. 27, where the drive route line is shown in white solid line. The UE velocity is 5km/h and the interfering sites are shown with white dotted lines. Along with drive route, spectrum efficiency (transmission rate vs. bandwidth) for CJT and single TRP transmission are tested and collected.

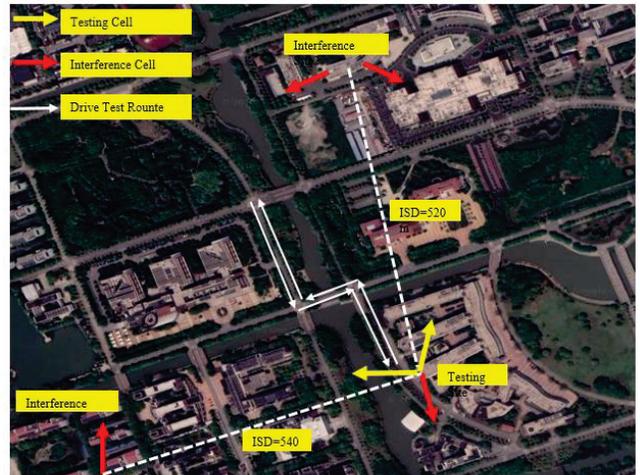

**Fig. 27.** Scenario of SU field test for CJT.

Figure 28 shows the CDFs of spectrum efficiency comparison of multi-TRP CJT and single TRP transmission along with drive route line. CJT achieve consistent UE experience with almost same spectrum efficiency along the drive route. Compared with single TRP transmission, more than 30% performance gain for cell edge (5%) could be obtained by CJT.

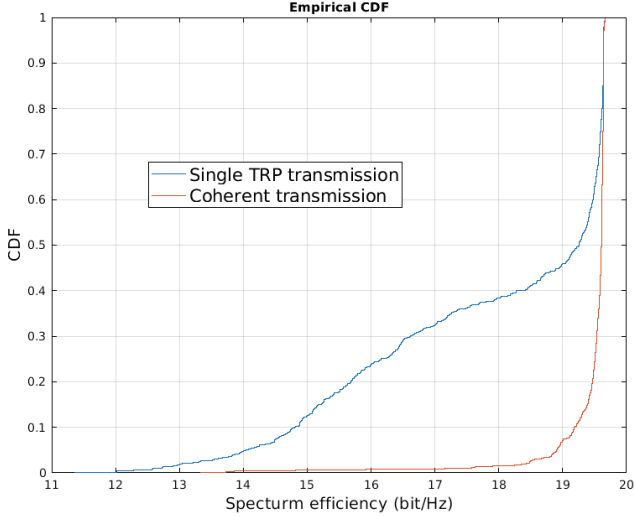

Fig. 28. Performance gain of CJT with SU.

We also test the multi-UE performance with CJT, and the multi-UE test scenario is shown in Fig 29. To verify the performance gain for different scenarios, 4 test cases are considered as shown in Fig. 30, where different UE locations with different RSRP gap are considered.

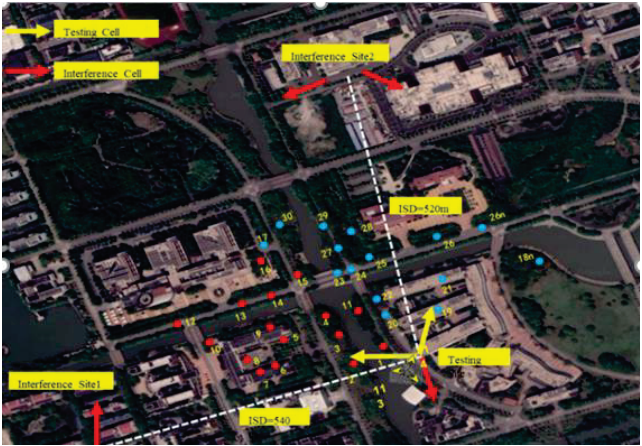

Fig. 29. Scenario for multi-UE field test for CJT.

As shown in Fig. 30, all the UE locations are divided into 2 group where one marked with red color, and one marked with blue color. During MU field test, one red UE location and one blue UE location are chosen for 2-UE test. For each group, four regions are defined:

- ✧ Region 1 means UE received RSRP difference from different TRPs is less than 3 dB;
- ✧ Region 2 means the RSRP difference is between 3 dB and 10 dB;
- ✧ Region 3 is with the RSRP difference between 10 dB and 15 dB;
- ✧ Region 4 means UE received RSRP difference from different TRP is more than 15 dB.

It can readily be predicted that the beneficial region is the region with small RSRP gap, where each TRP would have a big impact on the received signal at UE side.

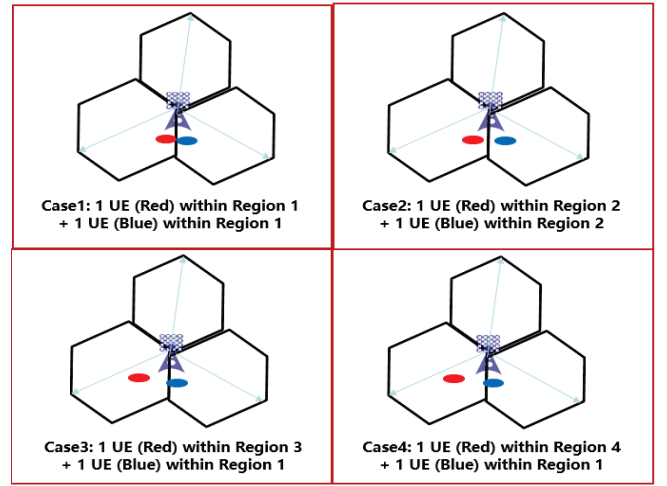

Fig. 30. Different cases with different UE distributions for multi-UE field trials.

The performance gain of CJT with multi-UE pairing for the 4 different cases is shown in Fig. 31. It is observed up to 98% gain can be achieved by CJT with 2 TRPs with MU. From Fig. 33, it also can be verified that UEs in the small RSRP difference region obtain more performance gain with CJT. In case 1, both UEs are in region 1 (RSRP gap < 3dB) and both TRP can have a big impact on the received signal for both UEs. For single TRP transmission, the SINR for each UE can be expressed as:

$$\text{SINR}_1 = \frac{\left\| H_1^{4\times 64} P_1^{64\times R_1} \right\|^2}{\left\| H_1^{4\times 64} P_2^{64\times R_2} \right\|^2 + n}, \quad (24)$$

$$\text{SINR}_2 = \frac{\left\| H_2^{4\times 64} P_2^{64\times R_2} \right\|^2}{\left\| H_2^{4\times 64} P_1^{64\times R_1} \right\|^2 + n}, \quad (25)$$

where data transmission for each UE suffers from big interference from another TRP. Note that $R_1$ and $R_2$ are the transmission layers for 2 UEs. For CJT transmission, the SINR for each UE can be expressed as:

$$\text{SINR}_1 = \frac{\left\| H_{CJT,1}^{4\times 128} P_1^{128\times R_1} \right\|^2}{\left\| H_{CJT,1}^{4\times 128} P_2^{128\times R_2} \right\|^2 + n}, \quad (26)$$

$$\text{SINR}_2 = \frac{\left\| H_{CJT,2}^{4\times 64} P_2^{128\times R_2} \right\|^2}{\left\| H_{CJT,2}^{4\times 64} P_1^{128\times R_1} \right\|^2 + n}. \quad (27)$$

With CJT, larger beamforming gains in $\left\| H_{CJT,1}^{4\times 128} P_1^{128\times R_1} \right\|^2$ and $\left\| H_{CJT,2}^{4\times 64} P_2^{128\times R_2} \right\|^2$ can be achieved. The interference from the data transmission for another UE in $\left\| H_{CJT,1}^{4\times 128} P_2^{128\times R_2} \right\|^2$ and $\left\| H_{CJT,2}^{4\times 64} P_1^{128\times R_1} \right\|^2$ can be suppressed significantly also. To this end, larger throughputs can be achieved by CJT.

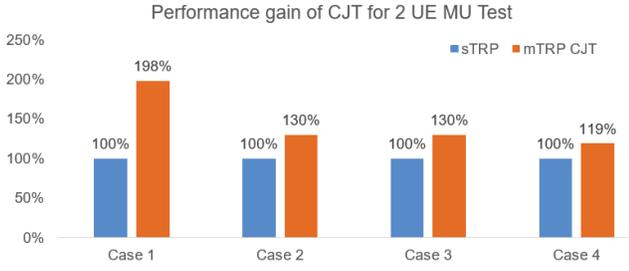

Fig. 31. Performance gain of CJT with multi-UE.

*C. Simulation Results for mmWave/FR-2 Bands*

In this section, the simulation results for the 3GPP Release 17 beam management enhancements are presented and analyzed. As described earlier, the motivation of the unified TCI framework introduced in 3GPP Release 17 for beam management is to reduce the latency and the overhead of beam indication, thereby enhancing system performance especially in high-speed scenarios. To quantify these benefits, we simulated and analyzed two scenarios, first is a high-speed train (HST) scenario, illustrated in Fig. 32 the details of which are described in [6], second is a dense urban highway (DUH) scenario illustrated in Fig. 33. In the highway scenario, the UE is moving between point P and point Q. The UE is moving at a speed of 120 kmph (33.3 m/sec), UE's throughput is sampled every 1 m (30 ms), with 100 sample points i.e., simulation duration is 3 seconds. Performance is evaluated for beam indication using DCI, with a latency of 0.5 ms and a BLER of 1%, and using MAC-CE, with a latency of 3 ms and a BLER of 10%.

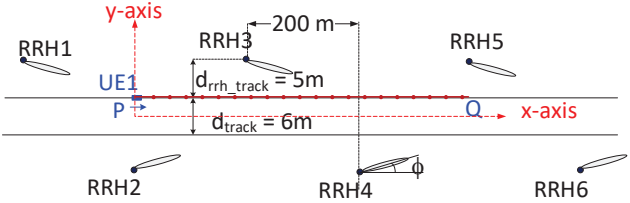

Fig. 32. HST scenario including UE trajectory [6].

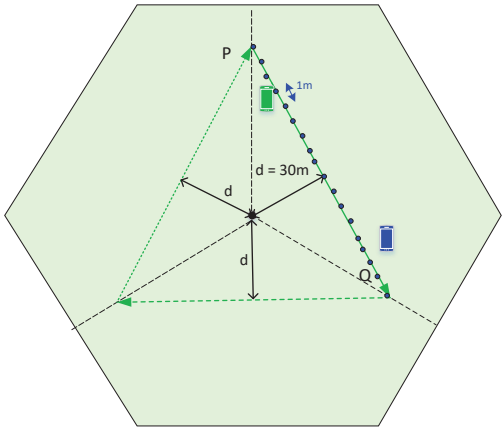

Fig. 33. Dense Urban Highway (DUH) Scenario, with UE simulated to be moving from point P to point Q.

For the HST scenario, Fig. 34 illustrates the benefit of using DCI with the unified TCI state framework over legacy (5G-NR Release 16) MAC CE for beam indication. As can be seen from the figure, on average, DCI using the unified TCI state framework provides a 10% spectral efficiency gain over MAC CE. The benefit using DCI for beam indication with low latency, is most apparent as the UE moves close to the RRH, when the rate of angular change and the according beam change is the highest. A longer latency in switching beams leads to a larger a drop in SINR and consequently lower performance. The channel carrying the beam indication is transmitted using the earlier beam, hence a drop in performance of this channel could lead to more hybrid automatic repeat request (HARQ) re-transmissions, i.e., even longer latency and could eventually lead to radio link failure. Figure 35 illustrates the impact of longer latency on the SINR of the channel carrying the beam indication as the UE moves past RRH B (RRH 3 in Fig.32).

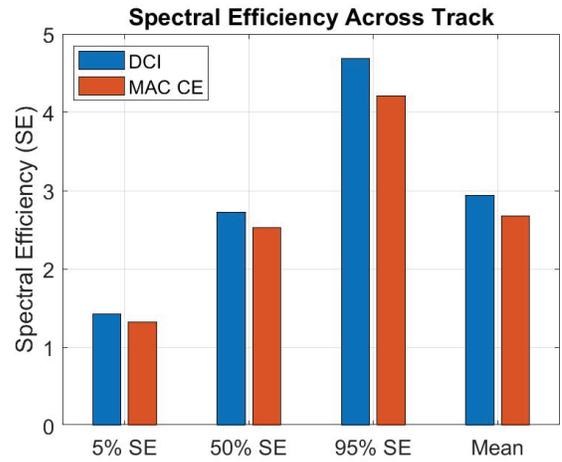

Fig. 34. Spectral efficiency for a UE moving from point P to point Q for HST scenario.

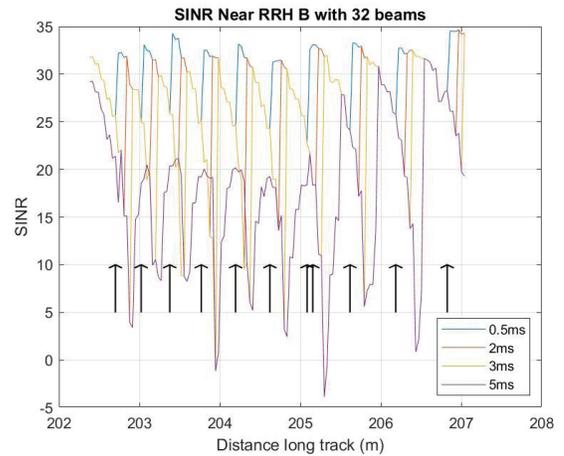

Fig. 35. Impact of beam latency on SINR of channel carrying beam indication signal. The arrows represent the points of beam change. RRH B is RRH 3 in Fig. 32.

For the DUH scenario, Fig. 36 illustrates the benefit of using DCI with the unified TCI state framework over legacy (NR Rel-16) MAC CE for beam indication. As can be seen from the figure, on average, DCI using the unified TCI state framework provides over 10% spectral efficiency (SE) gain over MAC CE. The rate of angular change and hence beam change is fastest as the UE moves past the nearest point to the gNB (BS). Hence, the benefit of using DCI for beam indication over MAC CE is greatest at the that point. This can be seen from Fig. 37, where the CDF plot of the normalized throughput shows a gain of 60% for DCI over MAC CE at the 50-th percentile point for nearest point to the gNB, while the gain drops to just 15% when we consider 3000 slots (0.75 second) around the nearest point to the gNB.

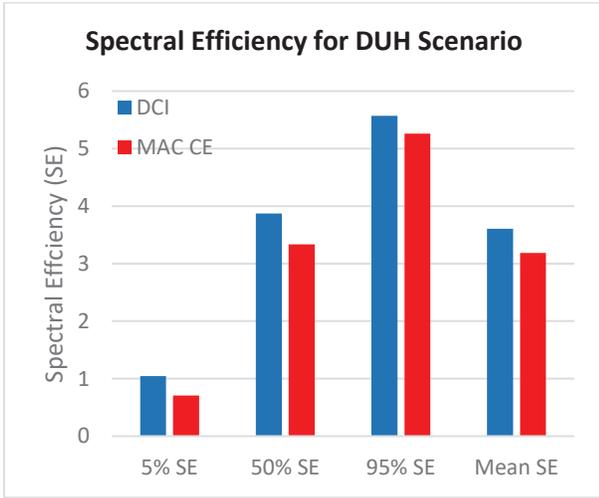

**Fig. 36.** Spectral efficiency for a UE moving from point P to point Q for DUH scenario.

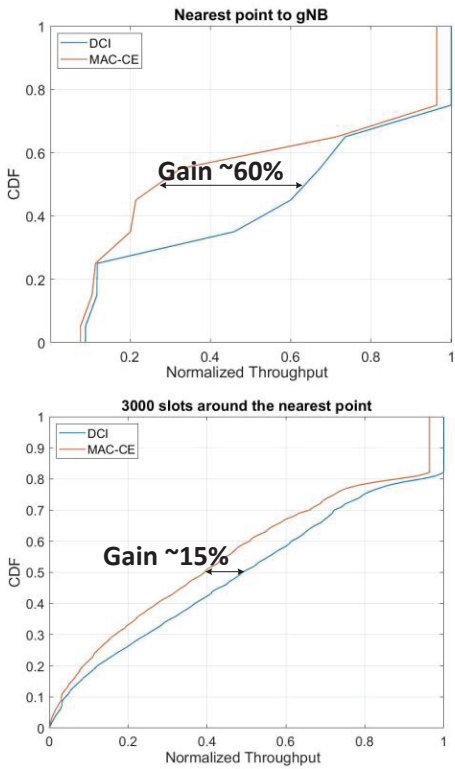

**Fig. 37.** CDF of normalized throughput around nearest point to the gNB (BS).

## VI. CONCLUSIONS

In this paper, we provided our considerations for massive MIMO Evolution towards 5G-Advanced in Release 18. To begin with, we present comprehensive review of massive MIMO evolution for TDD and FDD from earlier releases, i.e. from Releases 15 to 17. Starting from commercial massive MIMO architecture, fundamental characteristics like beamforming gain, beamwidth and angular coverage are achieved by a structure of antenna sub-arrays. CSI reporting and acquisition framework specified in earlier releases were further elaborated in the paper. Type-I single panel (for FR-1) and multiple panel (for FR-2) codebooks can provide low resolution CSI quantization for SU-MIMO. On the other hand, Type II codebook can provide high resolution CSI quantization typically used for multi-UE MIMO and was enhanced in multiple releases by taking advantage of frequency coherency or channel reciprocity to compress codebook parameter like combination coefficients. We explained fundamental SRS mechanisms like antenna switching and frequency hopping, considering implementation constraints of UE RF and power. For FR-2 specific, mmWave application has motivated multi-beam operation and beam maintenance via managing and indicating TCI state or SRS ID.

Starting from above overview and moving forward toward Release 18, we presented our considerations to address critical issues of NR system design:

- To support FDD CJT for multi-TRPs, we presented new codebook structure to support coherent reception at UE with precise CSI reporting and acquisition across TRPs. Starting from Type II codebooks, it is feasible to retrieve only sparse and dominant basis taking advantage of both spatial and frequency correlation from multi-TRPs jointly.
- To support TDD CJT for multi-TRPs, inter-cell SRS interference is problematic when using SRS for DL channel estimation. We demonstrated that with random CS hopping applied to SRS sequence transmission, inter-cell interference can be averaged and whitened so that targeted SRS sequence and associated MIMO channel can be estimated more precisely with time filtering.
- To support uplink massive MIMO for higher order UL MU-MIMO, we presented a design of increasing the number of orthogonal DMRS ports by simply doubling OCC length.
- To support better CSI acquisition in mobility, we presented new codebook structure to allow UE making CSI prediction based on historical CSI measurements. Assuming that Doppler frequency can be static within a time window, it is feasible to compress the latest and predicated CSI in a CSI reporting to reduce reporting overhead and achieve reasonable accuracy simultaneously.
- To support mmWave application for multi-TRPs, we proposed to extend unified TCI state framework to support more TCI states in DCI format to reduce beam switch latency.

SLS were demonstrated to support above physical layer optimization. It is clear that CJT is essential for 5G-advanced as it can provide profound performance gain with better codebook design and SRS CS hopping. System throughput for uplink massive MIMO with more orthogonal uplink DMRS ports can be roughly doubled. CSI prediction and associated CSI compression at UE can compensate inaccurate CSI for mobility. For FR-2 bands and high mobility, a unified TCI framework can provide fast beam switch by DCI so as to throughput gain, especially around those nearest points to the BS, for optimal beam tracking in mobility. Last but not least, we presented TDD field measurements in outdoor to demonstrate the feasibility and performance gain of CJT. Both single-UE and multi-UE MIMO were tested in field. Compared to single TRP based transmission, CJT can significantly improve UE experience at cell edge to achieve ubiquitous spectrum efficiency across network. For multi-UE MIMO in CJT, different MU pairing strategies were also compared. Whilst 3GPP Release 18 just has just begun as a Work Item, we believe this paper can provide useful and practical optimization techniques for the deployment of massive MIMO, which will be the foundation of future 3GPP RAN standards.

## APPENDIX

**Table A-1.** SLS Assumptions for CSI Enhancements. The *terminology* and *abbreviations* of all parameters is consistent with 3GPP specifications and can be found from [6,8-18,21,25] as well as the 3GPP references quoted within the table.

| Simulation Parameter | Value |
| --- | --- |
| Scenario | Urban Macro |
| Frequency Range | Intra-cell CJT scenario: <br> - FDD: 700 MHz, SCS=15kHz <br><br> Inter-cell CJT scenario: <br> - FDD: 2.1GHz, SCS=15kHz <br> - TDD: 3.5GHz, SCS=30kHz |
| Channel model | From 3GPP TR 38.901 |
| Antenna setup and port layouts at BS | Intra-cell CJT scenario: <br> - FDD: 8 ports (M, N, P, Mg, Ng; Mp, Np)=(2,2,2,1,1,2,2), <br> (dH, dV) = (0.5, 0.8). <br><br> Inter-cell CJT scenario: <br> - FDD: 32 ports: (M, N, P, Mg, Ng; Mp, Np) = (8,8,2,1,1,2,8), <br> - TDD: 64 ports, (M, N, P, Mg, Ng; Mp, Np) =(8,4,2,1,1,4,8). <br> (dH, dV) = (0.5, 0.8) |
| Antenna setup and port layouts at UE | 2RX: (M, N, P, Mg, Ng; Mp, Np) = (1,1,2,1,1,1,1), (dH,dV) = (0.5, 0.5)λ |
| BS power and height | 46dBm and 25m |
| UE antenna mode | From 3GPP TR 36.873 |
| Noise figure | 9 dB |
| Traffic mode | Non-full buffer mode |
| Modulation | Up to 256QAM |
| Channel estimation | Non-ideal channel estimation with LS operations |

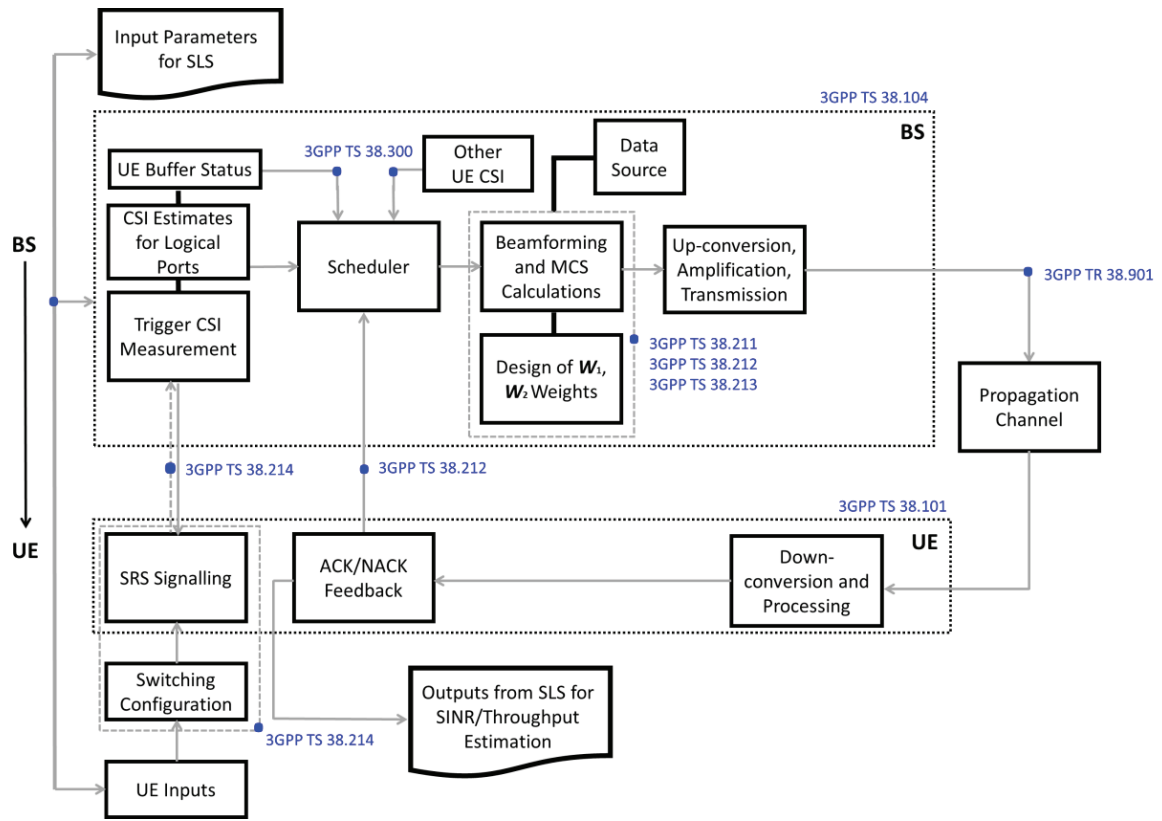

**Fig. A-1.** An example of a TDD downlink SLS following 3GPP standardized procedure for transmission from a BS to a UE. This is utilized in the results presented in Sec. V. A.

**Table A-3.** SLS assumptions for UL-MIMO. The *terminology* and *abbreviations* of all parameters is consistent with 3GPP specifications and can be found from [6,8-18,21,25].

| Simulation Parameter | Value |
| --- | --- |
| **Center Frequency** | 2.6 GHz |
| Scenario | IIOT, TRP number=18, UE number = 6 per TRP |
| **TRP antenna configuration** | (M, N, P, Mg, Ng; Mp, Np) = (1, 2, 2, 1, 1; 1, 2), (dH, dV) = (0.5, 0.8) |
| **UE antenna configuration** | (M, N, P, Mg, Ng; Mp, Np) = (1, 1, 2, 1, 1; 1, 1), (dH, dV) = (0.5, 0.5) |
| **MIMO scheme** | Multi-UE, multi-stream |
| **Numerology** | 14 OFDM symbol per slot, 30kHz sub-carrier spacing (SCS) |
| **UE speed** | 3 km/h |
| **Scheduling granularity** | 24 resource blocks (RB) |
| **DMRS** | Type 2 DMRS, double-symbol |
| **Channel estimation** | Non-ideal channel estimation with LS operations |
| **Traffic model** | Full buffer |
| **UE distribution** | 100% indoor |